\documentclass[11pt]{article} 

\usepackage{amsmath,amsthm,latexsym,amssymb,amsfonts,epsfig}

\newtheorem*{Theorem}{Theorem}

\newcommand{\be}{\begin{equation}}
\newcommand{\ee}{\end{equation}}
\newcommand{\ba}{\begin{eqnarray}}
\newcommand{\ea}{\end{eqnarray}}

\title{{\sf Symmetry reduction, gauge reduction, backreaction and}\\
{\sf consistent higher order perturbation theory}} 
\author{
{\sf T. Thiemann}$^1$\thanks{{\sf 
thomas.thiemann@gravity.fau.de}}\\
\\
{\sf $^1$ Institute for Quantum Gravity, FAU Erlangen -- N\"urnberg,}\\
{\sf Staudtstr. 7, 91058 Erlangen, Germany}\\
}
\date{{\small\sf \today}}

\makeatletter
\@addtoreset{equation}{section}
\makeatother

\begin{document} 

\maketitle

{\sf

\begin{abstract}
For interacting classical field theories such as general relativity 
exact solutions typically can only be found by imposing physically 
motivated (Killing) {\it symmetry} assumptions. Such highly symmetric 
solutions are then often used as {\it backgrounds} in a 
{\it perturbative} approach 
to more general non-symmetric solutions.

If the theory is in addition a {\it gauge} theory such as general relativity,   
the issue arises how to consistently combine the perturbative expansion 
with the gauge reduction. For instance it is not granted that the 
corresponding constraints expanded to a given order still close under 
Poisson brackets with respect to the non-symmetric degrees 
of freedom up to higher order. 

If one is interested in the problem of {\it backreaction} between 
symmetric and non-symmetric dgrees of freedom, then one also
must consider the symmetric degrees of freedom as dynamical variables
which supply additional terms in Poisson brackets with respect to 
the symmetric degrees of freedom and 
the just mentioned consistency problem becomes even more complicated.

In this paper we show for a general theory how to consistently combine 
all of these notions. The idea is to {\it first} perform the {\it exact} gauge 
reduction on the {\it full} phase space 
which results in the reduced phase space of observables 
and physical Hamiltonian respectively and {\it secondly} expand that 
physical Hamiltonian perturbatively. Surprisingly, 
this strategy is not only practically feasible but also avoids 
the above mentioned tensions.

There is also a variant of this strategy that employs only a {\it partial}
gauge reduction with respect to some of the non-symmetric degrees of freedom
on the {\it full} phase space. We show that in perturbation theory the 
left over constraints close up to higher orders but not exactly, unless 
there is only one of them such as in cosmology. Since such 
{\it classically anomalous} constraints are problematic to quantise, 
the {\it full} gauge reduction for which these issues are absent 
is preferred in this case. 
\end{abstract}

\section{Introduction}
\label{s1}

In interacting field theories (such as general relativity (GR)) to find 
the general, exact solution of the field equations is not feasible.
However, exact solutions can often be found when high amounts of 
(Killing) symmetries are imposed because this effectively reduces 
the number of dimensions and simplifies the associated PDE system. 
An extreme case are the cosmological solutions by asking for homogeneous 
spacetimes which reduces the Einstein equations to a system of ODE's.
Other well known examples are spherically symmetric or axi-symmetric 
spacetimes describing black holes of Schwarzschild or Kerr type respectively.
See \cite{Pirani} for an almost complete list of known exact solutions of 
GR with and without matter.

To find more general solutions one can use perturbative methods. One 
uses an exact symmetric solution as {\it background} and the 
non-symmetric deviations of the 
field from that symmetric background as a {\it perturbation}. The field
equations can then be expanded order by order with respect to these 
perturbations and one can attempt to find solutions to any desired 
accuracy with respect to the perturbative order. 

This simple idea gets complicated when the field theory under consideration 
is a gauge theory such as GR wich is subject to the spacetime diffeomorphism 
gauge group. We will work in the Hamiltonian setting with an eye towards 
canonical quantisation and discuss here only gauge symmetries 
generated by constraints of first class 
type in Dirac's classification \cite{Dirac} and only the case of a 
totally constrained system for which the Hamiltonian of the theory is a 
linear combination of constraints. Every gauge system can be reduced to 
that form by getting rid of the second class constraints replacing the
Poisson bracket by the  
Dirac bracket and by parametrising a possibly present physical Hamiltonian.

In the presence of such gauge symmetries, the exact 
field equations split into two sets, one of them presenting constraints 
on the initial data and the other one presenting dynamical equations. 
The dynamical equations in turn can be obtained from the Hamiltonian
in terms of Poisson brackets on the full phase space. For GR 
the constraints are known as spatial diffeomorphism and Hamiltonian 
constraints which are of first class. Systems of first class constraints 
by definition are in involution under the Poisson bracket on the 
full phase space and generate gauge motions on it. The constraint 
surface in the phase space is the set of points where the constraints vanish 
and the reduced phase space is the set of gauge orbits of the constraint 
surface which comes equipped with its own Poisson bracket. Functions on 
the reduced phase space are by construction gauge invariant and present 
so called Dirac observables.    
   
When it comes to perturbations around symmetric backgrounds, 
we split the phase space coordinates into two sets, one of them presenting 
the symmetric background, the other one the non-symmetric perturbations. 
One expands the constraints in powers of these perturbations and may 
call the constraints truncated to n-th order in those perturbations 
constraints of n-th order. At this point one has to make a decision whether 
one wants to take {\it backreaction} into account or not. If not, one treats 
the background variables as prescribed external functions and just 
considers the phase space described by the non-symmetric perturbations.
If yes, the background variables are still dynamical, we still consider 
the full phase space but use coordinates adapted to the (Killing) symmetry 
under investigation. In either case, it is not at all clear whether 
the n-th order constraints are in involution, at least up to higher 
than $n-$th order. Without backreaction this is is known to be the 
case for $n\le 2$ (see e.g. \cite{BDTT}) 
for a general theory explaining the success
of cosmological, Schwarzschild and Kerr perturbation theory resulting 
in the well known Mukhanov-Sasaki, Wheeler-Zerilli and Teukolsky equations 
respectively \cite{MS, WZ, Teukolsky}. 
With backreaction this is known not to be the 
case in examples 
but there is a procedure for how to correct this within perturbation 
theory and with backreaction 
for the case of cosmology and for $n=2$ \cite{Gomar}. 
Beyond $n=2$, to the 
best knowledge of the author, there is no commonly accepted procedure 
for how to reconcile perturbation theory with gauge symmetry (with or 
without backreaction), see e.g.
\cite{HOCPT,HOCPT1} and references therein for two concrete proposals 
within the Lagrangian and Hamiltonian formulation respectively which 
are not obviously equivalent. This 
poses a severe problem because without a consistent notion of 
gauge invariance, the observables of the theory cannot be extracted 
and the connection with phenomenology becomes veiled. Furthermore,
on the practical side, within the current proposals the gauge invariant
variables have to be recalculated every time one increases the 
perturbative order. It would be more convenient to disentangle 
gauge invariance from perturbation theory.         

A possible avenue has been suggested in \cite{GHTW} where one passed 
to the exact reduced phase space before performing perturbation theory.
This solves the tension because the gauge symmetry has now been 
taken care of to all orders. One can then extract the physical Hamiltonian
and observables
using deparametrisation and perform perturbation theory in the usual
way that is familiar from unconstrained theories, directly in terms of 
fully gauge invariant observables and including backreaction.
However, in \cite{GHTW} this could only be achieved by adding artificial 
dust matter to GR. While the effects of the dust were shown to be 
negligible in the late universe for small dust energy tensor, it would be
desirable to be able to follow the strategy of \cite{GHTW} also without 
dust matter.\\
\\
In this paper we consider the more difficult but physically more 
interesting case of including backreactions. In principle 
this is nothing but the full theory, however, written in 
field coordinates adapted to the symmetry of interest. We 
show, for a general 
first class theory subject to mild structural conditions that 
are motivated by the perturbative structure of the constraints of GR 
when using Killing symmetry reductions, that the machinery of 
deparametrisation of gauge systems \cite{Henneaux}
can be consistently combined with 
perturbation theory. This can be done 
in two versions. Either one performs a full reduction of all constraints 
or only a partial reduction corresponding to constraints that are naturally
associated with some of the non-symmetric degrees of freedom. The exact
full reduction immediately leads to a fully gauge invariant physical 
Hamiltonian and is analogous to \cite{GHTW} while the exact 
partial reduction 
is analogous to \cite{Gomar} and keeps some left over constraints 
which are in involution. Surprisingly,
these exact expressions, which are in general
known only implicitly, can be accessed perturbatively as we show 
in this paper. The corresponding perturbative techniques do not require more 
than standard Taylor expansions of the constraints and for GR these 
have been worked out to some extent also to higher than second order 
for several symmetric backgrounds. However, the consistency between 
gauge reduction and perturbation theory imposes that the bits and pieces
of these expressions be assembled in a novel way.

At second 
order, the resulting expressions reproduce the results 
of \cite{Gomar} for second order cosmological perturbation theory
with backreaction. Since our method 
is not confined to $n=2$, it embeds the method of \cite{Gomar} in a 
wider context. Our method can also be applied to the case of more than 
one unreduced constraint. In that case, the unreduced, perturbed constraints 
do close up to higher orders (trivially, in the case of just one unreduced 
constraint, they close exactly at any order) but not exactly. 
Therefore, as quantising constraints which are already classically anomalous 
are problematic in the quantum theory, the partial reduction version appears
to be disfavoured in this case (e.g. black hole perturbation theory). 
As an alternative, one may follow the theory developed
in \cite{Gambini} precisely for such ``only approximately'' first 
class constraints which is designed to deal with this problem, roughly 
by considering the closure violation as rendering the approximate constraints 
second class and applying the quantisation methods developed for second 
class constraints \cite{Henneaux}. \\
\\
\\
The architecture of this paper is as follows:\\
\\
In section two we explain the general research question
and motivate the structure 
of the canonical gauge system with respect to the symmetry reduction using 
the Killing reductions that one uses in GR. We pin point the problems that 
arise when one tries to combine standard Taylor expansions of constraints
with gauge reduction.

In section three we review the reduced phase space approach to first class 
gauge systems, in particular the notion of relational Dirac observables
subordinate to a choice of gauge fixing conditions or 
``clock'' functions and corresponding 
physical Hamiltonian. Equivalently one may use the gauge fixing 
approach leading to the so-called ``true degrees of freedom'' subordinate 
to such a choice of gauge fixing condition and corresponding ``reduced'' 
Hamiltonian.    

In section four we combine sections two and three which leads to a  
split of the canonical coordinates into four groups: symmetric or 
non-symmetric gauge degrees of freedom and symmetric or non-symmetric 
true degrees of freedom. Likewise, the constraints split into two groups 
corresponding to symmetric or non-symmetric smearing functions. 
In abuse of notation we call 
them symmetric or non-symmetric but note that in general they neither
close among themselves nor just generate gauge transformations just 
on the symmetric or non-symmetric degrees of freedom.   
We then solve the non-perturbative constraints 
for the gauge degrees of freedom exactly, subordinate to a choice of 
gauge fixings. The split described suggests to adapt the gauge fixings 
to that split and thus to reduce the symmetric or non-symmetric 
constraints with respect to some of the symmetric or 
non-symmetric gauge degrees 
of freeedom respectively. The result of that step is a reduced 
Hamiltonian which 
just depends on the true degrees of freedom, both symmetric and 
non-symmetric. \\
For sufficiently complicated systems, this 
exact reduced Hamiltonian will be known only implicitly involving inversions 
of functions etc. which makes it of little practical use. Surprisingly, 
it is still possible to derive a practically 
useful perturbation theory. This comes in two versions,
the fully reduced and the partially reduced version described 
above and we outline the full version in the same section.
We will see that this involves nothing but standard Taylor 
expansions of the constraints but assembelled in a novel fashion in order 
to meet the consistency with gauge invariance.
Hence standard unreduced perturbation 
theory is still valid, but our approach assembles its ingredients directly 
and unambiguously into objects that have a physical (gauge invariant) meaning.

In section five we lay out the 
reduction theory for the partially reduced version, also 
called ``reduction in stages''. The 
motivation for such an approach could be to perform a classical gauge 
reduction with respect to some of the non-symmetric degrees of freedom 
eliminating just the 
non-symmetric constraints and then to quantise the resulting remaining
symmetric constraints. A prominent example for this is the hybrid 
approach to quantum cosmology \cite{Hybrid}.  
Again, this is practically useful only when applying 
perturbation theory to the resulting constraints. 
The problem is that those perturbed resulting constraints do not close 
exactly but only up to higher order in perturbation theory which 
poses a challenge for quantisation, unless 
there is only one of them, which is the case in cosmology which is why 
\cite{Hybrid} is successful. We show that the ``in stages'' approach 
for systems with only one symmetric constraint reproduces the 
framework of \cite{Gomar} obtained for 2nd order cosmological perturbation
theory with backreaction and thus embeds this framework into our
perturbation theory that is also valid at higher orders.

In section six we detail the perturbation theory for the remaining 
constraints of a partially reduced system including backreaction. 
The formulae we write can directly be applied to higher order 
cosmological perturbation theory with backreaction and thus can be applied 
for instance in the study of cosmological non Gaussianities. 
We supply these formulae for general 
orders in terms of an iteration scheme and solve the scheme explicitly 
for {\it third order}.

In section seven we conclude and give an outlook into the many 
future applications of the present work, such as quantum 
black hole perturbation theory which we have treated by the methods 
presented in the present paper in \cite{BHPT}.

\section{Symmetry reduction, gauge reduction, backreaction and perturbation
theory} 
\label{s2}

In the first subsection  
we motivate the general structure of the gauge systems for which we intend 
to develop perturbation theory around symmetric configurations 
using the example of general relativity (GR) where it is physically motivated 
to consider Killing reductions. In the second subsection we abstract from 
the example of GR and summarise the structure that we found 
in the first subsection. This structure is crucial for 
the constructions that follow in the subsequent
sections. 

\subsection{Killing reductions in GR and mode decompositions}
\label{s2.1}

We consider a group G which acts via diffeomorphisms $\varphi_g,\; g\in$G
on tensor fields $T$ on the spacetime manifold $M$. We have 
$\varphi_g\circ \varphi_{g'}=\varphi_{gg'}$ and $\varphi_{1_{{\sf G}}}=
1_{{\sf Diff(M)}}$. If the set of tensor fields under 
consideration includes a metric then it is sufficient to consider purely
co-variant tensor fields so that the action is just by pull-back
$T\mapsto \varphi^\ast T$ otherwise we also have to consider 
push-forwards for mixed tensors carrying also contra-variant structure.
For the purpose of this motivation we confine ourselves to theories with 
metric fields, the additional details needed for the general case are easy 
to supply.
   
A tensor field is called symmetric with respect to G iff 
$\varphi_g^\ast T=T$ for all $g\in$G, otherwise non-symmetric. We also 
call a symmetric tensor a {\it zero mode} for reasons that become clear 
shortly. The group G acts of course also on $M$ via $\varphi_g$ and 
$M$ has invariant submanifolds $N=\varphi_g(N)$ for all $g\in$G. 
We assume that the manifold $M$ has the product structure 
$M=M_1 \times M_2$ with corresponding coordinates $(\rho,\theta)$ such 
that $\rho$ labels the invariant submanifolds $N=N_\rho,\;\rho=$const. 
and $\theta$ are ``angular'' coordinates on the $N_\rho$. In particular, the 
functions $\rho$ are invariant scalars on $M$ and $\varphi_g$ acts 
non-trivially only on the coordinates $\theta$. In other words, $M$ is 
foliated by the leaves $N_\rho$ diffeomorphic to $M_2$ with foliation 
parameters $\rho$.

This warped structure 
motivates to perform harmonic analysis on $M_2$: We note that a (pseudo-)
tensor 
$t$ on $M_2$ has both a ``spin'' transformation of its indices and an
``orbital'' transformation of its arguments. 
The spin part can be considered 
as a finite dimensional representation $\pi_s$ of G and the 
orbital part as an infinite
dimensional one $\pi_o\equiv \phi_\cdot$ and thus the tensor transfoms in the 
infinite 
dimensional tensor product 
representation $\pi_s\otimes \pi_o$ as 
\be \label{2.1}
[\varphi_g^\ast t](\theta)=\pi_s(g)\cdot t(\varphi_g(\theta))
\ee
Suppose that such a representation
can be decomposed into irreducibles $\pi$. An 
{\it irreducible tensor harmonic} 
$t_\pi$ 
on $M_2$ of type $\pi$ where $\pi$ is an 
irreducible representation of G corresponds precisely to such a decomposition. 

A distinguished role in what follows 
is played by an invariant metric field $\Omega$ on $M_2$ of Euclidian 
signature
i.e. $\varphi_g^\ast \Omega=\Omega,\; g\in$G which we assume to exist. 
It is a tensor harmonic of degree two with respect to the trivial 
representation. 
It follows that $d\mu(\theta)=|\det(\Omega)(\theta)|^{1/2}\;d\theta$ 
is an invariant measure on $M_2$, that is  
\be \label{2.2}
\int_{M_2} \; d\mu(\theta)\; f(\varphi_g(\theta))=
\int_{M_2} \; d\mu(\theta)\; f(\theta)=:\mu(f)
\ee
for all measurable functions $f$ on $M_2$. We assume that G and therefore 
$M_2$ is compact and $\mu$ or $\Omega$ can therefore be normalised such that 
$\mu(1)=1$. Here $d\theta$ is the Lebesgue measure on $M_2$. In the 
non-compact case one ususally compactifies $M_2$ (e.g. toroidally) and 
considers a decompactification limit at the end of the quantisation 
process (thermodynamic limit).

Using $\Omega$ we can define an inner product on the space 
$L^d_2$ of square integrable tensors on $M_2$ of degree $d$
\be \label{2.3}
<t,t'>:=\int_{M_2} \; d\mu(\theta)\;
\overline{t(\theta)} \cdot 
[\otimes^d\;\Omega^{-1}(\theta)] \cdot
t'(\theta)
\ee
where $\cdot$ denotes 
the only possible contraction of indices of the tensors involved 
with the inverse metric $\Omega^{-1}$ which grants that the inner 
product is positive. The complex conjugation in (\ref{2.3}) can of course 
be dropped when the tensors are real valued but often it is convenient 
to consider complex valued tensors. We note that G acts unitarily on 
$L^d_2$. 

To become more concrete, we consider first degree $d=0$, i.e. scalars.
Then ${\cal H}=L_2(M_2,d\mu)$ can be decomposed into invariant subspaces 
${\cal H}_\pi$ corresponding to irreducible representations $\pi$. We 
pick an ONB $L^\pi_k,\; k=1,..,\dim(\pi)$ wrt (\ref{2.3}) in ${\cal H}_\pi$  
and consider the map $I_{\pi',\pi};\;{\cal H}_\pi\;\to 
{\cal H}_{\pi'}$ defined by $I_{\pi,\pi'}\cdot v:=\sum_{k'}
L^{\pi'}_{k'}\;<L^{\pi'}_{k'},v_1>_{{\cal H}}$. Using unitarity we 
see that $I_{\pi,\pi'}$ is an intertwiner. By definition of irreducible 
representations, the intertwiner must be trivial, i.e. the identity when 
the representations are equivalent and zero when they are not (Schur's 
lemma). It follows that the $L^\pi_k$ form an orthonormal system 
of ${\cal H}$ and we assume that it is in fact an orthonormal basis i.e. 
that Plancherel's theorem holds. This decomposition denotes the ``orbital''
part $\pi_o=\pi$ 
of the decomposition. Now tensors of degree $d>0$ can be obtained 
by combinations of 
multiple actions of the $\Omega$ compatible torsion free covariant 
differential $D$ on scalars together with contractions by $\Omega$ and 
$\eta$ where $\eta$ is the totally skew pseudo-tensor on $M_2$. 
A tensor is called polar and axial respectively if the number of $\eta$
factors used is even and odd respectively so that the tensor itself is 
a tensor and pseudo tensor respectively. Given a basis of (pseudo-) tensors of 
degree $d$ so obtained from scalars, we can decompose them into orthogonal 
subspaces with respect to the fibre metric $\otimes^d \Omega^{-1}$. 
These spaces are also invariant under G because they are built from covariant
tensor operations. This decomposition denotes the ``spin'' part $\pi_s$ 
of the decomposition. Now given a tensor of a certain spin type we can 
restrict the scalars on which the invariant tensor operations act to 
the orbital space labelled by $\pi_s$. This completes the concrete 
description of irreducible tensor harmonics of type $\pi_s\otimes \pi_o$.  

The relevance of this with regard to perturbation theory is now as 
follows: Given a tensor field $T$ on $M=M_1\times M_2$, for 
fixed $\rho$ we can consider it as a collection of tensor fields on $M_2$.
To do this we split the coordinates $x^\mu,\;\mu=0,..,m=m_1+m_2-1$ into 
$\rho^\alpha,\; \alpha=0,..,m_1-1$ and $\theta^A,\;A=1,..,m_2$ and 
accordingly each tensor $T$ on $M$ of total degree $d$ can be considered 
as a collection of tensors of degree $d_2=0,1,..,d$ on $M_2$ with fixed 
$d_1=d,d-1,..,0$ indices taking values in the set of values of $\alpha$ 
and the remaining 
the values of $A$. Each tensor in this collection 
transforms in some finite dimensional representation of G. This 
is because for
G compact, its irreducible representations are completely 
classified, they are all finite dimensional and and every representation 
decomposes into 
irreducibles \cite{Hall}. Therefore, all tensor fields $T$ on $M$ can be 
uniquely decomposed into irreducible tensor harmonics. The expansion 
coefficients in that decomposition are tensors on $M_1$ i.e. we have 
a neat decomposition of any $T$ on $M$ into a linear combination 
of tensor products $t^\pi_1\otimes t_\pi$ where $t_\pi$ is an
irreducible tensor harmonic of type $\pi$ on $M_2$, called a {\it mode} 
and $t_1^\pi$ is some tensor on $M_1$. The perturbative structure 
is now defined by distinguishing between the tensors $t_1^\pi$ corresponding 
to the trivial representation, i.e. the zero modes or symmetric tensors,
and the tensors $t_1^\pi$ corresponding to non-trivial representation, i.e.
the non-zero modes or non-symmetric tensors. The zero modes are 
declared as background degrees of freedom while the non zero-modes are 
declared as perturbations.

So far we have used spacetime language. In the canonical approach one 
considers globally hyperbolic spacetimes $(M,g)$ whose underlying manifold 
is diffeomorphic to a Cartesian product $\mathbb{R}\times \sigma$ for 
some $m-1$ manifold $\sigma$ \cite{Sanchez}. We will assume that      
the $\varphi_g$ preserve the time axis and therefore all that we have said 
applies also to $\sigma=\sigma_1\times M_2$ with 
$M_1=\mathbb{R}\times \sigma_1$. Next to the tensor fields themselves
in the canonical formulation now one also needs 
their first time derivatives on the Cauchy surface $\sigma$ or equivalently    
their conjugate momenta which are tensor densities of weight one of dual
type (i.e. purely contra-variant if the tensor fields are purely co-variant).
Thus for each tensor mode $t_1^\pi$ we have a momentum $p_1^\pi$. These 
are automatically canonically conjugate. To see this, we consider the 
symplectic structure at fixed $t$ 
\be \label{2.4}
\Theta=\int_\sigma\; d^{m-1}x\; P\cdot [\delta T]
\ee
where $P$ is conjugate to $T$ and and both $P,T$ have not yet been decomposed 
into tensor harmonics. Using coordinates  such that
$d^{m-1} x=d\rho \; d\mu$, we now 
expand both $P,T$ into tensor harmonics 
$P^\pi\otimes t_\pi , T^\pi\otimes t_\pi$ and see that the mixed 
terms drop out because (\ref{2.4}) is of the form of the inner product 
(\ref{2.3}), specifically
\be \label{2.5}
\Theta=\sum_\pi\; \int_{\sigma_1}\; d\rho_1\; P^\pi\cdot [\delta T^\pi]
\ee
where $\rho=(t,\rho_1)$ was used.

Next we come to the constraints. By construction, the constraints 
$C=C(T,P)$ are 
themselves tensor fields on $\sigma$ of density weight one constructed 
from the canonical fields $T,P$ while their 
smearing functions $f$ are dual tensors of density weight zero. We decompose 
$f$ into tensor harmonics $f^\pi\otimes t_\pi$, do the same with 
$P,T$ and integrate out $M_2$ resulting in 
\be \label{2.6}
C(F)=\int_\sigma\; d^{m-1}x\; f\cdot C=\sum_\pi \;\int_{\sigma_1} \; 
d\rho_1\;
f^\pi\cdot <t_\pi,C>=:\sum_\pi \int_{\sigma_1} \; d\rho_1\; f^\pi\cdot C_\pi 
\ee
Note that $C_\pi$ can depend non-trivially on all degrees of freedom 
$T^{\pi'}, P^{\pi'}$.

Perturbation theory now consists in singling out the symmetric degrees of 
freedom $Q_B=T^{\pi_t},P_B=P^{\pi_t}$ with trivial representation $\pi_t$ 
as background variables and to keep those 
$Q_\pi=T^{\pi},P_\pi=P^{\pi}$
for non trivial representation
$\pi\not=\pi_t$ as first order perturbation. Note that for each $\pi$ we can 
have several ``species'' that may result from additional matter content 
or because there are several ways to couple spin and orbital dependence 
into the same $\pi$ or because of the remaining tensor structure on $M_1$.
We suppress a corresponding species index in order not to clutter 
the notation. We can perform a Taylor 
expansion of the $C_\pi$ in terms of the $Q_{\pi'},\; P_{\pi'}$ 
with coefficients that depend only on the $Q_B,P_B$. 
Let $C_{\pi,(n)}$ be the n-th order perturbation of $C_\pi$ 
i.e. a homogenous polynomial in the $Q_{\pi'},P_{\pi'}$ 
of degree $n$. Then an important 
observation is that $C_{\pi,(0)}=0$ 
if $\pi\not=\pi_t$ and $C_{\pi,(1)}=0$ for $\pi=\pi_t$
because of the orthogonality properties of (\ref{2.3}). 

Finally it will be convenient to assume that the tensorial type with respect 
to $M_2$ of the smearing 
functions $f$ of the constraints also appears among the list of canonical 
tensor fields or their canonical momenta, possibly after performing a 
canonical transformation. In GR these tensorial types are scalar and vectorial
and the assumption just made is always met when decomposing the spatial 
metric with respect to the tensor type on $M_2$.\\ 
\\
We close this subsection by mentioning how the above general theory 
fits with the well known examples relevant for cosmology and black holes
in four spacetime dimensions:\\
i.\\
For cosmology we assume that $\sigma$ is compact, say a 3-torus (choosing 
the torus sufficiently large we cannot observationally distinguish it 
from $\mathbb{R}^3$) whence G=SO(2)$^3$ is the translation group of the torus.
The tensor harmonics are then just the Fourier modes of the torus 
labelled by a vector $\pi\in \mathbb{Z}^3$ and 
the manifold $\sigma_1$ is zero dimensional so that the integral 
over $\rho_1$ in (\ref{2.6}) is discarded. The zero modes with $\pi=0$
in this case 
are just homogeneous metric and matter degrees of freedom.\\
ii.\\
For spherically symmetric Schwarzschild black holes the relevant group
is G=SO(3) and the tensor harmonics are the well known spherical tensor 
harmonics \cite{SphTensorHarm} which are labelled by indices 
$\pi=(I,l,m)$ where 
$I$ is a discrete index depending on the tensor degree and $l,m$ are the usual
quantum numbers labelling the scalar harmonics $Y_{l,m}$ familiar from 
the theory of angular momentum. The zero modes are functions that depend 
only on the radial variable $\rho_1=r$.\\
iii.\\
For axi-symmetric Schwarzschild black holes the relevant group
is G=SO(2) and the tensor harmonics are again just the Fourier modes of the 
1-torus labelled by $\pi\in \mathbb{Z}$. 
The zero modes are functions that depend only on the radial and axial variable
$\rho_1=(r,z)$.\\
\\
Finally notice that while we have here only treated the case of bosonic 
tensor fields, an extension of tensor harmonics to spin harmonics 
i.e. fermionic fields is easily possible by passing to the covering 
group of the above groups which is still compact 
and otherwise performing the same decomposition into irreducibles, see 
e.g. \cite{SpinHarmonics}.

\subsection{Summary of the symmetry, gauge and perturbative 
structure and notation}
\label{s2.2}

We consider a phase space with canonically conjugate 
coordinates $k^{{\cal A}},i_{{\cal A}}$ where in field theory 
$\cal A$ takes values in a countably infinite index set (the mode labels
$\pi$, the species labels and if $\sigma_1$ is not zero dimensional 
further labels e.g. corresponding to an orthonormal basis of 
$L_2(\sigma_1,d\rho_1)$) but in what follows that range could also be 
finite. Likewise we have non-perturbative constraints $C_\mu$ where again 
$\mu$ has has countably inifinite range in field theory but we can also 
consider finite range in what follows. We invoke the information about the 
symmetry structure of the background by splitting $\mu$ into pairs 
$\mu=(a,j)$ where $a$ labels the zero (symmetric) modes of the smearing 
function $f^\mu$ and $j$ the non-zero (non-symmetric) modes. Accordingly 
we have symmetric and non-symmetric constraints $C_a,\; C_j$ respectively as 
coefficients of $f^a, f^j$ respectively.      

As motivated at the end of the previous subsection we can perform a 
corresponding 
split also among the canonical coordinates. But unless the theory is 
topological 
there will be additional degrees of freedom. Accordingly we split
\be \label{2.7}
(k^{{\cal A}},i_{{\cal A}})=
((q^a,p_a),\;(x^j,y_j),\;(Q^A,P_A),\;(X^J,Y_J))
\ee
where the meaning is as follows: Both pairs $(q,p)$ and $(Q,P)$ 
correspond to zero modes, i.e. they are symmetric degrees of freedom.
Both pairs $(x,y)$ and $(X,Y)$ 
correspond to non-zero modes, i.e. they are non-symmetric degrees 
of freedom. This emphasises their grouping with respect to the symmetry 
and perturbative structure, that is, one will expand with respect to 
$x,y,X,Y$ around $q,p,Q,P$. On the other hand we can group them with respect
to the gauge structure: The index picture suggests to consider 
the pairs $(p,q)$ and $(x,y)$ as pure gauge degrees of freedom while 
the pairs $(P,Q)$ and $(X,Y)$ are considered as true degrees of freedom.

In other words we have a twisting of four sectors corresponding to 
the symmetry and gauge aspect: There are both symmetric and non-symmetric 
observables $(P,Q),(X,Y)$ and both symmetric and non-symmetric 
gauge variables $(p,q),(x,y)$. The fact that these four pairs of variables
are conjugate within the respective pair and have vanishing Poisson brackets 
between 
variables of different pairs was motivated in (\ref{2.5}) above.

We can now develop perturbation theory on the unreduced phase space 
coordinatised by (\ref{2.7}), that is, we perform a Taylor expansion 
of the constraints 
\be \label{2.8}
C_a((p,q),(P,Q),(x,y),(X,Y)),\; 
C_j((p,q),(P,Q),(x,y),(X,Y))
\ee
with respect to $x,y,X,Y$ at fixed $p,q,P,Q$. We denote by $C_{a(n)}$ 
the $n$-th order contribution to $C_a$ in that expansion which is a 
homogneous polynomial of degree $n$ in $x,y,X,Y$ with coefficients 
which may depend non-polynomially on $p,q,P,Q$. The meaning of 
$C_{j(n)}$ is similar. On the other hand the expansion of the 
constraints to n-th order is denoted as 
\be \label{2.9}
C_a^{(n)}=\sum_{k=0}^n\; C_{a(k)},\;
C_j^{(n)}=\sum_{k=0}^n\; C_{j(n)}
\ee
and we write $O(n)$ for any function on phase space whose perturbative 
expansion contains homogeneous orders of degree $n$ or higher.

The observation made towards the end of the previous subsection translates 
into the statement that 
\be \label{2.10}
C_{a(1)}=0,\;C_{j(0)}=0
\ee

By assumption, the non-perturbative constraints are in involution, that is 
there are structure functions $\kappa$ on the full phase space such that 
\ba \label{2.11}
\{C_a,C_b\} &=& 
\kappa_{ab}\;^c\; C_c +\kappa_{ab}\;^j\; C_j
\nonumber\\
\{C_a,C_j\} &=& 
\kappa_{aj}\;^b\; C_b +\kappa_{aj}\;^k\; C_k
\nonumber\\
\{C_j,C_k\} &=& 
\kappa_{jk}\;^a\; C_a +\kappa_{jk}\;^l\; C_l
\ea
where the Poisson brackets $\{.,.\}$ 
are with respect to all phase space coordinates
and the indices $a,b,c,...$ and $j,k,l,...$ have the same range
respectively, summation 
over repeated indices being implied.

We introduce a corresponding perturbative notation 
$\kappa_{\ast\ast(n)}\;^{\ast},\;
\kappa_{\ast\ast}\;^{\ast(n)}$ for the structure functions and introduce 
the following symmetric and non-symmetric Poisson bracket respectively
\be \label{2.12}
\{F,G\}_S:=
\{F,q^a\}\;\{p_a,G\}+\{F,Q^A\}\;\{P_A,G\}
-\{G,q^a\}\;\{p_a,F\}-\{G,Q^A\}\;\{P_A,F\},\;
\{F,G\}_{\bar{S}}:=\{F,G\}-\{F,G\}_S
\ee
which just takes derivatives with respect to the symmetric and non-symmetric 
degrees of freedom respectively. It follows that $\{O(m),O(n)\}_S=O(m+n)$
and $\{O(m),O(n)\}_{\bar{S}}=O(m+n-2)$ which has the advantage that we can 
better keep track of the perturbative order. We can use this 
and matching of perturbative order 
to derive the infinite hierachy of exact relations for $N=0,1,..$ 
\be \label{2.13}  
\sum_{m+n=N;\;m,n\ge 0}\; \{C_{(m)},C_{(n)}\}_S+
\sum_{m+n=N+2;\;m,n\ge 1}\; \{C_{(m)},C_{(n)}\}_{\bar{S}}
=\sum_{m+n=N;\;m,n\ge 0}\; \kappa_{(m)}\; C_{(n)}
\ee
where we have suppressed the indices $a,b,c,j,k,l$ which are the same as in 
(\ref{2.11}). 

The system (\ref{2.13}) together with (\ref{2.10}) contains important 
information, we only state more explicitly the first few of them relevant 
for perturbation theory up ton $n=2$: \\
$N=0$: \\
i.\\ 
Since $\{C_a(0),.\}_{\bar{S}}=0=C_{a(1)}$
\be \label{2.14}
\{C_{a(0)},C_{b(0)}\}=\kappa_{ab(0)}\;^c\; C_{c(0)} 
\ee
i.e. the $C_{a(0)}$ are in involution with respect to the full Poisson 
bracket.\\
ii.\\ 
Since $C_{j(0)}=0$ 
\be \label{2.15}
\{C_{j(1)},C_{k(1)}\}_{\bar{S}}=\kappa_{jk(0)}\;^a\; C_{a(0)} 
\ee
i.e. the $C_{j(1)}$ close modulo $C_{a(0)}$ but only with respect to the 
non-symmetric bracket.\\
iii.\\ 
Since $C_{a(1)}=C_{j(0)}=0$
\be \label{2.16z}
\kappa_{aj(0)}\;^b=0
\ee
$N=1$: \\
i.\\ Using $C_{a(1)}=0$
\be \label{2.16b}
0=\kappa_{ab(1)}\;^k\; C_{c(0)} 
+\kappa_{ab(0)}\;^j\; C_{j(1)} 
\ee
ii. \\
Using $C_{a(1)}=C_{j(0)}=0$
\be \label{2.16a}
\{C_{a(0)},C_{j(1)}\}_S+\{C_{a(2)},C_{j(1)}\}_{\bar{S}}
=\kappa_{aj(0)}\;^k C_{k(1)}+\kappa_{aj(1)}\;^b C_{b(0)}
\ee
$N=2:$ \\
Using $C_{a(1)}=0=C_{j(0)}$
\be \label{2.17}
\{C_{a(0)},C_{b(2)}\}_S+\{C_{a(2)},C_{b(0)}\}_S
+\{C_{a(2)},C_{b(2)}\}_{\bar{S}}
=\kappa_{ab(0)}\;^c\; C_{c(2)} 
+\kappa_{ab(0)}\;^j\; C_{j(2)}
+\kappa_{ab(1)}\;^j\; C_{j(1)} 
+\kappa_{ab(2)}\;^c\; C_{c(0)}
\ee
We may now apply the following logic: As we consider the non-symmetric 
degrees of freedom as perturbations and since in GR the smearing functions 
also are dynamical degrees of freedom (lapse and shift functions, they are 
variables conjugate to the primary constraints of GR) in the expansion 
of the Hamiltonian 
\be \label{2.18}  
H(f,g)=f^a C_a + g^j C_j
\ee
we should consider also $f^a$ as of zeroth order and $g^j$ of first order. 
This argument is substantiated by another one which considers the 
gauge transformation $\delta F=\{H(f,g),F\}$ generated by (\ref{2.18}).
For a symmetric degree of freedom $F\in \{q,p,Q,P\}$ 
which is of zeroth order we have $\Delta F=\{H(f,g),F\}_S$ while 
for a non-symmetric degree of freedom $F\in \{x,y,X,Y\}$ which is 
of first order we have $\Delta F=\{H(f,g),F\}_{\bar{S}}$. Since 
$C_a=C_{a(0)}+O(2)$ while $C_j=C_{j(1)}+O(2)$ the zeroth order character 
of $F$ is preserved by $\delta F$ iff $f$ is considered as zeroth order 
and the first order character 
of $F$ is preserved by $\delta F$ iff $g$ is considered as first order. 

Following this logic the $N-$th order truncation of $H$ is given by
\be \label{2.19}
H^{(N)}(f,g)=f^a\; C_a^{(N)}+g^j C^{(N-1)}_j
\ee
and we may ask whether the N-th order 
truncations $f^a C_a^{(N)}, g^j C_j^{(N-1)}$ 
are in involution 
for various values of $N$, either exactly or up to $O(N+1)$ terms at least.   
Due to $C_{a(1)}=C_{j(0)}=0$ we have $H^{(0)}(f)=f^a C_{a(0)}$ 
which {\it is} in involution as demonstrated by (\ref{2.14}) both with 
respect to the full bracket and the symmetric bracket only. 
This says that symmetry reduction excluding perturbations produces a
consistent gauge system. 

For $N=2$ we find 
\be \label{2.20}
H^{(2)}(f)=f^a\;[C_{a(0)}+C_{a(2)}]+g^j C_{j(1)}=:C^{(2)}(f)+C^{(1)}(g)
\ee
The analysis of the closure of the constraints now depends on whether one 
wants to 
incorporate backreaction or not. With backreaction, the system should 
close with respect to the full bracket $\{.,.\}$, without it, it should 
close with respect to the non-symmetric bracket $\{.,.\}_{\bar{S}}$. \\
\\
I. No backreaction. \\
From (\ref{2.15}) we have directly 
\be \label{2.22}
\{C^{(1)}(g),C^{(1)}(g')\}_{\bar{S}}=
g^j (g')^k
\kappa_{jk(0)}\;^a C_{a(0)}
=g^j (g')^k
\kappa_{jk(0)}\;^a C_a^{(2)}-\{g^j (g')^k
\kappa_{jk(0)}\;^a C_{a(2)}\}
\ee
We can close the algebra of the $C^{(1)}(g)$ 
exactly if we impose on $p,q,P,Q$  that 
$C_{a(0)}=0$ 
which without backreaction is possible as the background variables
are considered as an external structure.
This is what is usually done \cite{Langlois}
and (\ref{2.22}) it is then in fact Abelian.
Otherwise it closes up to a term $C^{(2)}(f)$ modulo an O(4) correction 
displayed as the curly bracket term in (\ref{2.22})
if $g,g'$ count as first order each. Next from (\ref{2.16a})
\be \label{2.23}
\{C^{(1)}(g),C^{(2)}(f)\}_{\bar{S}}=
\{C^{(1)}(g),C_{(2)}(f)\}_{\bar{S}}
=g^j f^a [\kappa_{ja(0)}\;^k C_{k(1)}+\kappa_{ja(1)}\;^b C_{b(0)}
-\{C_{j(1)},C_{a(0)}\}_S]
\ee
where we used that $C_{a(0)}$ has vanishing 
$\{.,.\}_{\bar{S}}$ brackets. We can close (\ref{2.23}) exactly 
if again we impose on $p,q,P,Q$ that $C_{a(0)}=0$ there and that it is 
a common critical point of the $C_{a(0)}$ i.e. their Hamiltonian 
vector fields vanish there so that the subtracted term in (\ref{2.23}) 
vanishes. Together with the fact that $C_{a(1)}=0$ this means that 
the Hamiltonian vector field of the non-perturbative $C_a$ vanishes 
in the full phase space 
at the chosen $p,q,P,Q$ and at $x=y=X=Y=0$ which then is a point on 
the full constraint surface. When $a$ takes only one value (cosmology)
this is again standard \cite{Langlois}. Since the subtracted term in 
(\ref{2.23}) is O(2) this is in fact the only option. Finally
from (\ref{2.17})
\ba \label{2.24}
&& \{C^{(2)}(f),C^{(2)}(f')\}_{\bar{S}}=
\{C_{(2)}(f),C_{(2)}(f')\}_{\bar{S}}
\\
&=& f^a (f')^b[
\kappa_{ab(0)}\;^c\; C_{c(2)} 
+\kappa_{ab(0)}\;^j\; C_{j(2)}
+\kappa_{ab(1)}\;^j\; C_{j(1)} 
+\kappa_{ab(2)}\;^c\; C_{c(0)}]
-\{C_{a(0)},C_{b(2)}\}_S-\{C_{a(2)},C_{b(0)}\}_S
\nonumber\\
&=& f^a (f')^b[
\kappa_{ab(0)}\;^b\; C_c^{(2)} 
+\kappa_{ab(0)}\;^j\; C_{j(2)}
+\kappa_{ab(1)}\;^j\; C_{j(1)} 
+(\kappa_{ab(2)}\;^c - \kappa_{ab(0)}\;^c)\; C_{c(0)}]
\nonumber\\
&& -\{C_{a(0)},C_{b(2)}\}_S-\{C_{a(2)},C_{b(0)}\}_S
\nonumber
\ea
which closes exactly if $C_{a(0)}$ and its Hamilonian vector field vanishes 
at $q,p,Q,P$ and if in addition $\kappa_{ab(0)}\;^j=0$ there. From 
(\ref{2.16z}) which holds everywhere in the full phase space,
this is indeed the case as the $C_{j(1)}$ are linearly independent.\\
\\
II. Including backreaction.\\   
Using the above results we find the full and exact Poisson bracket relations
\ba \label{2.25} 
&& \{C^{(1)}(g),C^{(1)}(g')\}=
\{C^{(1)}(g),C^{(1)}(g')\}_{\bar{S}}+
\{C^{(1)}(g),C^{(1)}(g')\}_S
=g^j (g')^k\kappa_{jk(0}\;^a C_{a(0)}
+\{C^{(1)}(g),C^{(1)}(g')\}_S
\nonumber\\
&=& g^j (g')^k\kappa_{jk(0}\;^a C_a^{(2)}
+\{\{C^{(1)}(g),C^{(1)}(g')\}_S-
g^j (g')^k\kappa_{jk(0}\;^a\; C_{a(2)}\}
\nonumber\\
&& \{C^{(1)}(g),C^{(2)}(f)\}=
\{C^{(1)}(g),C^{(2)}(f)\}_{\bar{S}}
+\{C^{(1)}(g),C^{(2)}(f)\}_S
\nonumber\\
&=&\{C_{(1)}(g),C_{(2)}(f)\}_{\bar{S}}
+\{C_{(1)}(g),C_{(0)}(f)\}_S
+\{C_{(1)}(g),C_{(2)}(f)\}_S
\nonumber\\
&=& g^j f^a
[\kappa_{ja(0)}\;^k C_{k(1)}+\kappa_{ja(1)}\;^b C_{b(0)}]
+\{C_{(1)}(g),C_{(2)}(f)\}_S
\nonumber\\
&=& g^j f^a
[\kappa_{ja(0)}\;^k C_{k(1)}+\kappa_{ja(1)}\;^b C_b^{(2)}]
+\{\{C_{(1)}(g),C_{(2)}(f)\}_S- g^j f^a \kappa_{ja(1)}\;^b C_{b(2)}\}
\nonumber\\
&& \{C^{(2)}(f),C^{(2)}(f')\}=
\{C^{(2)}(f),C^{(2)}(f')\}_{\bar{S}}
\{C^{(2)}(f),C^{(2)}(f')\}_S
=\{C_{(2)}(f),C_{(2)}(f')\}_{\bar{S}}
+\{C^{(2)}(f),C^{(2)}(f')\}_S
\nonumber\\
&=& [\{C_{(2)}(f),C_{(2)}(f')\}_{\bar{S}}
+\{C_{(0)}(f),C_{(2)}(f')\}_S
+\{C_{(2)}(f),C_{(0)}(f')\}_S]
+\{C_{(0)}(f),C_{(0)}(f')\}_S
+\{C_{(2)}(f),C_{(2)}(f')\}_S
\nonumber\\
&=& f^a (f')^b\;[
\kappa_{ab(0)}\;^c C_{c(2)}
+\kappa_{ab(0)}\;^j C_{j(2)}
+\kappa_{ab(1)}\;^j C_{j(1)}
\kappa_{ab(2)}\;^c C_{c(0)}
+\kappa_{ab(0)}\;^c C_{c(0)}]
+\{C_{(2)}(f),C_{(2)}(f')\}_S
\nonumber\\
&=& f^a (f')^b\;[
\kappa_{ab(0)}\;^c C_c^{(2)}
+\kappa_{ab(0)}\;^j C_{j(2)}
+\kappa_{ab(1)}\;^j C_{j(1)}
\kappa_{ab(2)}\;^c C_{c(0)}]
+\{C_{(2)}(f),C_{(2)}(f')\}_S
\nonumber\\
&=& f^a (f')^b\;[
(\kappa_{ab(0)}\;^c+\kappa_{ab(2)}\;^c) C_c^{(2)}
+\kappa_{ab(0)}\;^j C_{j(2)}
+\kappa_{ab(1)}\;^j C_{j(1)}]
+\{\{C_{(2)}(f),C_{(2)}(f')\}_S-
\kappa_{ab(2)}\;^c C_{c(2)}\}
\ea
where we used several times that $C_{a(0)}$ has vanishing 
$\{.,.\}_{\bar{S}}$ brackets. If we consider $g, g'$ as first oder 
objects, then the curly bracket term in the above relations is O(4) and 
can be ignored in 2nd order perturbation theory. Then all three equations 
close up to the last equation to second order provided that in addition
$\kappa_{ab(0)}\;^j=0$ everywhere. In view of (\ref{2.16z}) which holds 
everywhere in phase space and the linear independence and of $C_{a(0)},
C_{j(1)}$ this may actually hold but it is not entirely conclusive, 
so we must impose this as an extra condition if we are to follow 
this approach. It holds trivially if $a$ takes only 
one value (cosmology). 

If on the other hand we want (\ref{2.25}) to 
close exactly and not only up to higher order, then as it stands the system 
is inconsistent. This explains why considerable effort must be invested 
to achieve closure in cosmology \cite{Gomar}. The motivation to have 
exact closure is that while in the classical theory one may be able 
to control errors when working with approximate equations, in the quantum 
theory it is vital to have exact relations because the quantum constraints 
determine the size of the physical Hilbert space. Anomalous, i.e. not 
exactly closing constraints, over-constrain the system and yield a physical 
Hilbert space which is too small to allow for the correct classical limit.
The way that closure of (\ref{2.25}) is achieved in \cite{Gomar} is 
by a clever combination of i. the exact relation (\ref{2.15}) combined 
with ii. canonical transformations on the full phase correct up 
to second order and iii. absorption of higher order terms into smearing 
the functions $f^a$. 

However, it is clear that the method of \cite{Gomar}
repairs the system (\ref{2.25}) only if $a$ takes only one value (cosmology) 
and for $N=2$ only. In fact, one can easily generalise (\ref{2.25}) to higher 
orders but obtaining exact closure fails more and more severly the higher
the order. Also closure up to higher order only occurs if one considers $g,g'$
as of first order and one may feel uneasy with that viewpoint.
The methods of this paper extend the results of 
\cite{Gomar} 
to a general gauge system subject to the assumptions spelled out in 
section \ref{s2.1} for an arbitrary range of $a$ and for arbitrary $N$.
We follow a route very different from the route chosen 
in \cite{Gomar}. Yet, for $N=2$ and $a$ taking only one value,
we agree with \cite{Gomar}.

\section{Gauge reduction, gauge conditions, relational Dirac observables,
physical Hamiltonian}
\label{s3}

This section is for the benefit of the rader not familiar with gauge reduction
and the relational formalism. We explain here just the bare bones of this 
theory, see \cite{Henneaux} for all the details and \cite{TTReduced}
for an exposition closer in notation to the present section. 
Familiar readers may 
immediately jump to the next section.\\
\\
In this section we ignore the split with respect to the symmetric and 
non-symmetric variables and we ignore perturbation theory altogether
and just assume that the phase space is coordinatised
by canonical pairs $(r^\alpha,s_\alpha)$ and $(u^\mu,v_\mu)$ respectively 
where $(r,s)$ plays the role of $(Q,P),(X,Y))$ and $(u,v)$ the role 
of $((q,p),(x,y))$. The system is subject to first class constraints 
$C_\mu$. The above split of the phase space variables is assumed to be 
judiciously chosen in such a way that one may solve the constraints 
$C_\mu(r,s,u,v)=0$ (locally) in terms of the momenta $v_\mu$, that is 
we pass to equivalent constraints 
\be \label{3.1}
\hat{C}_\mu(r,s,u,v)=v_\mu+h_\mu(r,s,u) 
\ee
The dependence of the $C_\mu$ on the $v_\nu$ is usually polynomial and if 
the polynomial degree is higher than one then the functions $h_\mu$ are 
not unique but depend on finitely many ``branches''. We assume that 
some physical motivation (e.g. positivity of energy) 
has been given to select one of those branches and consider henceforth 
the corresponding sector of the phase space that contains the constraint 
surface defined by (\ref{3.1}).

Since the $\hat{C}_\mu$ are just linear combinations of the $C_\mu$ 
(with complicated phase space dependent coefficients) as 
$C=0 \Leftrightarrow \hat{C}=0$ 
on the sector considered, and the $C_\mu$ are first class, so must be 
the $\hat{C}_\mu$ i.e. there are structure functions $\hat{\kappa}$ such that 
\be \label{3.2}
\{\hat{C}_\mu,\hat{C}_\nu\}=\hat{\kappa}_{\mu\nu}\;^\rho\; \hat{C}_\rho
\ee
Since the $\hat{C}$ contains $v$ linearly, the l.h.s. of 
(\ref{3.2}) is independen of $v$. Thus for any $(r,s,u)$ we may
evaluate the l.h.s. by setting $v=-h(r,s,u)$ on the r.h.s.
As this vanishes identically, we conclude that the constraints 
$\hat{C}_\mu$ are {\it Abelian} on the full phase space 
$\{\hat{C}_\mu,\hat{C}_\nu\}=0$. This fact will be very handy in what follows.

We split the following exposition in the ``gauge invariance'' and the 
``reduction'' viewpoint. Both are equivalent descriptions of the 
reduced phase space. The gauge invariance viewpoint explicitly constructs 
the relational Dirac observables as functions on the full phase space.
Here ``relational'' refers to the fact that one must select a set of 
gauge fixing conditions that enter explicitly into the construction:
One can ``project'' any phase space function into a Dirac observable 
relative to these gauge fixing conditions, but neither the constraints 
nor the gauge fixing conditions are installed. By contrast, the 
reduction viewpoint explicitly installs both the constraints and the 
gauge fixing conditions. The degrees left over are called ``true degrees
of freedom''. The two viewpoints are equivalent in the sense 
that the relational Dirac observables and the true degrees of freedom 
subordinate to the same gauge fixing conditions are Poisson isomorpic
and the corresponding physical respectively reduced Hamiltonians 
generate isomorphic equations of motion on the reduced phase space.

\subsection{Gauge invariance viewpoint}
\label{s3.1}

Let $\tau^\mu$ be any parameters (``multi-fingered time''), 
let $G^\mu:=\tau^\mu-u^\mu$ be {\it gauge fixing conditions} for the 
``clock'' variables $u$ and let $F=F(r,s,u,v)$ be any function of the 
full phase. Define the ``relational observable'' 
\be \label{3.3}
O_F(\tau):=F+\sum_{n=1}^\infty\; 
\frac{1}{n!}\; G^{\mu_1}..G^{\mu_n}\;
[X_{\mu_1}\cdot ..X_{\mu_n}\cdot F]
\ee
where $X_\mu\cdot F:=\{\hat{C}_\mu,F\}$ is the Hamiltonian vector field 
of $\hat{C}_\mu$. Then the functions (\ref{3.3}) share the following 
properties:\\
1.\\
They are Dirac observables $X_\mu\cdot O_F(\tau)=0$ for all $\mu,\tau$.\\
2.\\
Linearity $O_F(\tau)+O_{F'}(\tau)=O_{F+F'}(\tau)$.\\
3.\\
Compatibility with the pointwise product 
$O_F(\tau)\;O_{F'}(\tau)=O_{F\;F'}(\tau)$\\
4.\\ 
Compatibility with complex conjugation 
$\overline{O_F(\tau)}=O_{\bar{F}}(\tau)$ if the $G,C$ are real valued.\\ 
5.\\
Dirac bracket homomorphism 
\be \label{3.4}
\{O_F(\tau),O_{F'}(\tau)\}= 
\{O_F(\tau),O_{F'}(\tau)\}^\ast=
O_{\{F,F'\}^\ast}(\tau)
\ee
where the Dirac bracket is defined by (note that in our convention 
$\{\hat{C}_\mu,G^\nu\}=-\delta_\mu^\nu$)
\be \label{3.5}
\{F,F'\}^\ast=\{F,F'\}
-[\{F,\hat{C}_\mu\}\;\{G^\mu,F'\} 
-\{F',\hat{C}_\mu\}\;\{G^\mu,F\}]
\ee 
6. Let $t\mapsto \tau(t)$ be a 1-parameter curve, $F$ a function depending 
only on $r,s$ and $O_F(t)):=O_F(\tau(t))$. Then 
\be \label{3.6}
\frac{d}{dt} \; O_F(t)=\{H(t),O_F(t)\},\;\;
H(t)=O_{\dot{\tau}^\mu(t)\; h_\mu}(\tau(t))
\ee
i.e. $H(t)$ defines the generator of time evolution defined by the 
one parameter curve in clock space. 

Due to property 3. above, it is not necessary that one is able to compute 
the series involved in  (\ref{3.3}): Given a function 
$F=F(r,s,u,v)$ we may rewrite it as $\hat{F}(r,s,u,\hat{C})$ and find 
setting $R^\alpha(t):=O_{r^\alpha}(t),\;S_\alpha(t):=O_{s_\alpha}(t)$ 
that $O_F(t)=\hat{F}(R(t),S(t),\tau(t),\hat{C})$. Thus functions depending 
non-trivially on $u,v$ are of no interest when we pass to the constraint 
surface $\hat{C}=0$ which is preseved by definition of first class 
constraints. Therefore we are interested only the algebra of Dirac 
observables which are functions of $R,S$ only. In particular
\be \label{3.7}
H(t)=\dot{\tau}^\mu(t)\;h_\mu(R(t),S(t),\tau(t))
\ee
which is in general explicitly time dependent through the dependence of 
$h_\mu$ on $u$.

We note that the functions 
$R,S$ stay conjugate variables 
\be \label{3.8}
\{S_\alpha(t),R^\beta(t)\}=\delta_\alpha^\beta,\;      
\{S_\alpha(t),S_\beta(t)\}
=\{R^\alpha(t),R^\beta(t)\}=0
\ee

\subsection{Reduction viewpoint}
\label{s3.2}

A Poisson algebraically equivalent description can be given as follows:
We identify $v$ as the constrained momenta, $u$ as the pure gauge degrees 
of freedom and $r,s$ as the true degrees of freedom coordinatising 
the reduced phase space. Indeed the gauge cut $G=0$ through the 
constraint surface $\hat{C}=0$ is in one to one correpondence with 
the gauge orbits of points on $\hat{C}$ if i. the gauge $G=0$ 
can always be installed and ii. every gauge orbit is intersected precisely 
once (admissable gauge choice).
In order that the 
gauge conditions $G^\mu=\tau^\mu(t)-u^\mu$ be preserved under gauge 
transformations we must have $\dot{\tau}^\mu=\{\hat{f}^\nu\hat{C}_\nu,u^\mu\}$ 
that is $\hat{f}^\mu=\dot{\tau}^\mu=:\hat{f}^\mu_\ast$. 
The reduced Hamiltonian is then   
defined as the effective Hamiltonian on the reduced phase space whose 
action coincides with that of the original constrained Hamiltonian at the 
gauge cut.
Thus let $F=F(r,s)$ just depend on the true degrees of freedom $r,s$. Then
by definition $H=H(r,s;t)$ is the function obeying
\be \label{3.9a}
\{H(t),F\}=\{C(f),F\}_{\hat{C}=0,G=0,f=f_\ast}
\ee
or by rewriting the constraints $C(f)=\hat{C}(\hat{f})$ in terms of new 
Lagrange multipliers (which then depend on phase space but this immaterial 
as long as the linear map between $\hat{C}$ and $C$ is non-singular as 
both $f,\hat{f}$ become fixed as definite phase space functions by stability)  
\be \label{3.9}
\{H(t),F\}=\{\hat{C}(\hat{f}),F\}_{\hat{C}=0,G=0,\hat{f}=\hat{f}_\ast}
\ee
which is solved by  
\be \label{3.10}
H(t)=\dot{\tau}^\mu(t)\; h_\mu(r,s,\tau(t))
\ee
and thus coincides with (\ref{3.7}) under the Poisson bracket isomorphism 
$(r,s)\mapsto R,S$. \\
\\
Thus, given the extra structure provided by the gauge fixing conditions,
one may arrive at a reduced phase space description equivalently in terms 
of manifestly gauge invariant Dirac observables or the true degrees of 
freedom and given a one parameter family of gauge fixings one recovers a 
physical (i.e. observable) notion of dynamics even for totally constrained 
systems. 

We will apply the true degrees of freedom description to 
our gauge system combined with  perturbation theory with respect to
{\it non-symmetric observables}.

\subsection{Choice of gauge}
\label{s3.3}

What the above formalism does not tell is, which gauge fixing condition to
choose. Although all admissable choices are by definition related by a gauge 
transformation and thus merely correspond to different but gauge equivalent 
sections of the constraint surface, thus providing an explicit 
coordinatisation of the reduced phase space (space of gauge orbits),
the choice has a strong impact on the form of the reduced Hamiltonian 
as it decides which degrees of freedom to freeze and which are to be evolved.
As the variables that are prescribed define a reference frame (through 
the Lagrange multipliers that are fixed) it is clear that the physical 
Hamiltonian depends on that reference frame. Thus the selection of
gauge requires physical input, e.g. one may require
that the physical Hamiltonian acquires the form of the standard 
model Hamiltonian on Minkowski space in regions of the phase space that 
correspond to the flat Minkowski metric.  

Another issue is the following: In the previous sections we have 
considered gauge conditions of the form $q^a=\tau^a(t),\;x^j=\rho^j(t)$ 
which depend on a time parameter $t$ and thus define a one parameter
family of sections that gives rise to an evolution. Such an explicit 
time dependence of the gauge condition is a viable choice. However,
one would like to minimise such freedom when physical principles are 
one's disposal that help to downsize the number of possibilities. 
In the context of 
perturbation theory it is motivated to consider $\rho^j\equiv 0$ 
as the $x,y$ are ``small'' but 
as far as $\tau^a$ is concerned, this is not the case. When the index 
$a$ has univalent range (cosmology) then the choice of the remaining function
$\tau^a(t)$ is nothing but a reparametrisation of the time variable and thus 
essentially unique. However, when $a$ takes more than one value, in particular 
an infinite number, the 
issue becomes more serious because an infinite number of arbitrary parameters 
cannot be fixed by any experiment (predictivity).

Fortunately, precisely in situations when 
$a$ takes an infinite range there is another effect that in fact 
allows the time dependence of $\tau^a(t)$ to be trivial. Such gauge 
conditions are called ``coordinate conditions'' \cite{HanRegTei}. 
Typically this happens when our discrete description descends from a field 
theory rather than a finite dimensional Hamiltonian system. Note that even 
in this case and as mentioned at the beginning of section \ref{s2}, 
a discrete description is still possible by expanding the fields in 
terms of a countable set of mode functions $L_a$ (say Hermite functions if 
the non-compact part of the spatial topology consists of copies of 
$\mathbb{R}$). 

The origin of the effect is that 
the continuum constraints depend on spatial derivatives and that there is a 
boundary \cite{HanRegTei}. In that case, the symplectic structure 
and the constraints are no longer automatically 
finite and functionally differentiable. In order to ensure this, one 
must specify boundary or decay conditions on the canonical variables and the 
test functions that smear the constraints as one approaches the boundary
and the constraints can typically be made functionally differentiable
if one adds a boundary term which is finite when the specified 
decay behaviour holds but whose variation is singular and cancels 
the singular variation of the constraint. 
Thus the Hamiltonian in its functionally 
differentiable form and in terms of the original constraints 
not solved for $v_\mu=-h_\mu$ is now given by 
\be \label{3.10a}
H(f)=C(f)+B(f)
\ee
where $B(f)$ is the boundary term. A gauge transformation is defined by 
those $f$ for which $B(f)=0$ and a symmetry transformation by those 
$f$ for which $B(f)\not=0$. In both cases $f$ must obey the specified 
decay behaviour. The reduced phase space is still 
defined by coordinate conditions 
$G^\mu=u^\mu-\tau^\mu=0,\; u^\mu=u^\mu_\ast:=\tau^\mu$ 
with constant functions on the phase space $\tau^\mu$
and by the 
solutions of the constraints $C_\mu=0$. Since there are differential 
equations involved, while this system can be rewritten in the 
form $\hat{C}_\mu=v_\mu+h_\mu(r,s;u;V)=0,\;v_\mu^\ast:=-h_\mu(r,s;u=u_\ast;V)$
it depends on integration constants $V_A$.
These can be considered as initial conditions on the boundary when solving 
the differential equations, thus the constraints on the boundary are 
identically satisfied for all values of $V_A$. Therefore
the number of independent constraints is reduced by the number of those 
integration constants and we impose only as many gauge fixing conditions
as there are independent constraints. We may without loss of generality 
pick $V_A$ as the the values of $v_A$ on the boundary 
where $A$ runs through 
some index set denoting boundary degrees of freedom, it is contained 
in the bulk index set with labels $\mu$. We may then pick $U^A$
as the values of $u^A$ on the boundary 
and impose gauge fixing conditions only on $u^\mu$ where 
$\mu$ is not a boundary label and the gauge fixing condition is still
a constant function $\tau^\mu$ for those fiexed $u^\mu$ 
on the phase spce. This will be made more explicit below. 

The smearing functions are specified by gauge stability
\be \label{3.10b}
\{H(f),G^\mu\}_{u=u_\ast,v^\ast}=
\{H(f),u^\mu\}_{u=u_\ast,v^\ast}=0
\ee
because $\tau^\mu$ does not explicitly depend on time. However, 
because of the derivatives involved, which is why 
(\ref{3.10b}) is not imposed for 
all $\mu$, namely not for $\mu=A$, this now does not imply $f=0$ but 
yields a differential equation with non-trivial solutions 
$f^\mu=f^\mu_\ast(r,s;V,\lambda)$ which 
depends on yet other integrations constants $\lambda^A$. 
This $f_\ast$ obeys 
the decay behaviour specified but it generically corresponds to a symmetry
transformation and not a gauge transformation. Now for any function 
$F=F(r,s;U,V)$ we define the reduced Hamiltonian as before
\be \label{3.10c}
\{H,F\}=\{H(f),F\}_{w=w_\ast,f=f_\ast}
\ee
where $w=(u,v)$
but now it is less clear how to write $H$ explicitly. Since the boundary 
term is linear in $f$ we have (it does not depend on derivatives of $f$ 
after integrating by parts) 
\be \label{3.10d}        
B(f)=f^A \; j_A
\ee
where the index $\alpha$ runs through the boundary subset of the bulk 
index set to which $\mu$ belongs and the current 
$j_A$ defines the smearing function independent part of the boundary  
function on the full phase space. That $\lambda^A, U^A, V_A,
f^A$ are all labelled by the same index set is no coincidence but 
again 
due to the fact that we can define initial conditions for the differential 
equations to be solved, equivalent to intgration constants, at the boundary.  

Let $j^\ast=j_{w=w_\ast}$ then 
\ba \label{3.10e}
&& \{B(f)_{w=w_\ast},F\}_{f=f_\ast} = f^A_\ast \{j^\ast_A,F\}
\nonumber\\
&=& \{H(f)_{w=w_\ast},F\}_{f=f_\ast}
\nonumber\\
&=& \{H(f),F\}_{w=w_\ast,f=f_\ast}
+\{H(f),u^\mu\}_{w=w_\ast,f=f_\ast}\; \{v_\mu^\ast, F\}
-\{H(f),v_\mu\}_{w=w_\ast,f=f_\ast}\; \{u^\mu_\ast, F\}
\nonumber\\
&=& \{H,F\}
\ea
where in the second step we used that $H=B$ on the constraint surface,
in the third we expanded the Poisson bracket into explicit dependence 
of $H_{w_\ast}(f)$ on $r,s$ and implicit one through $w_\ast$ and in the 
last we noted that $u_\ast$ is a constant on the phase space and 
(\ref{3.10b}), (\ref{3.10c}). 

We see that a closed expression for 
$H$ can be obtained provided that there exists a function $\chi(j)$ such 
that $f^A_\ast=[\partial \chi/\partial j_A]_{j=j_\ast}$. Then
$H=\chi(j_\ast)$. One may use the freedom in the choice of the integration 
constants $\lambda$ on which $f^\ast$ depends in order to achieve that this 
condition holds because there will be typically as many integration constants,
i.e. initial conditions on the boundary, as there are boundary indices 
$\alpha$. The simplest case is that $f_\ast$ is a constant on the phase space,
then simply $H=f^A_\ast j_\alpha^\ast$.\\
\\
The above discussion follows closely \cite{HanRegTei} but deviates somewhat 
from the description in the previous section in that we have shifted the 
focus from the constraints $\hat{C}$ solved for the momenta 
to the original ones. 
We will therefore now describe an equivalent procedure which uses the 
$\hat{C}$ more directly and which offers a complementary point of view 
of how the integration constants $U,V,\lambda$ come into play.
We go back to the split description $\hat{C}_a, \hat{C}_j$ and  
for simplicity consider the case  
that the $\hat{C}_j$ have already been reduced by the 
gauge $x^j=0$ which is possible as in the form $\hat{C}_a,\hat{C}_j$ the 
constraints are mutually commuting. 
The case of a joint treatment of  
$q^a,x^j$ is similar. 

Since 
spatial derivatives acting on the mode functions can be expressed as 
a finite linear combination of mode functions again, a derivative in the 
continuum description translates into a kind of difference (with respect
to the label $a$) in the discrete description. This means that when 
computing Poisson brackets, one has to integrate (sum) by parts, leading 
to boundary contributions and bulk contributions of the bracket provided 
that there is a boundary as we will assume.
To describe 
this more explicitly, we assume (a situation often encountered in practice)
that 
we can split the label $a$ into pairs $(r,i)$ where 
partial integration (summing) is with respect to $r$ at fixed $i$ and 
w.l.g. $r$ has range in $\mathbb{R}^+_0$ ($\mathbb{Z}^+_0$).
The boundary contribution depends on the values of the fields 
$(q^a,p_a,R),\; R=(Q^A,P_A,X^J,Y_J)$ and the Lagrange multipliers $f^a$
at $r=\infty$ while the bulk contribution depends in particular 
on derivatives $f^{a\prime}$ (differences) of the $f^a$.  

Likewise, rewriting the constraints $C_a$ in the form 
$\hat{C}_a=p_a+h_a(q,R)$ there is an issue: As derivatives (differences) 
of the momenta $p_a$ are involved, to solve $C_a=0$ for the $p_a$ 
means solving a differential (difference) equation. As such equations 
require initial conditions or integration constants 
in order for a unique solution to exist,
in this case we should rather write $h_a(q,R)=-S_a(c_0;q,R)$ where 
$c_0=\{c_{r=0,i}\}$ 
is a collective notation for the integration constants. If we do not 
want to give $c_0$ the status of new degrees of freedom, we can solve 
the equation $p_{r=0,i}=S_{r=0,i}(c_0,q,R)$ for $c_0$ and insert the 
solution into 
$S_a(c_0,q,R)$ yielding $S_a(p_0,q,R)$ where now by construction 
$S_{r=0,i}(p_0,q,R)=p_{r=0,i}$. It follows that the constraints 
$\hat{C}_{0,i}:=\hat{C}_{r=0,i}\equiv 0$ 
are identically satisfied and thus we should in fact only impose 
gauge fixing conditions on $q^{r,i},\; r>0$. In this way the function
$h_a$ acquires the form $-S_a(p_0,q_0,\{q^b\}_{b>0},R)=:
\hat{h}_a(\{q^b\}_{b>0},\hat{R})$ where we have augmented the true degrees of 
freedom $R$ to $\hat{R}$ by the canonical pairs $(q^{0,i},p_{0,i})$
and $b=(r,i)>0\;\;\Leftrightarrow \;\; r>0$. The $q^{0,i}, p_{0,i}$ play the 
role of the $U^A, V_A$ above and we have relabelled the 
indices $A$ by $i$ as here we are dealing with split form of the 
constraints and consider only the $C_a$ and not all of the $C_\mu$.  

We can now rewrite the constraints $C_a$ in terms of the $\hat{C}_a$ but since
$C_a$ depends on derivatives (differences) of $p$ while 
in $\hat{C}_a$ no derivatives (differences) of $p$ appear, 
this rewriting necessarily involves derivatives (differences) of the 
$\hat{C}_a$. In the case that $C_a$ depends polynomially on the momenta 
$p$ and its derivatives (differences) and if only first derivatives 
(differences) are involved, $C_a$ can be written as a linear combination of  
$\hat{C}$ 
and its first derivatives $D\hat{C}$ (differences) with phase space dependent 
coefficients (simply by expanding $p_a=\hat{C}_a-\hat{h}_a$) say 
\be \label{3.11}
C_a=\sum_{b\ge 0} \; \gamma_a^b\; \hat{C}_b+\delta_a^b\; (D\hat{C})_b
\ee
Since vanishing of constraints means vanishing of their derivatives 
(differences) the vanishing of $\hat{C}_a$ for all $a\ge 0$ implies 
the vanishing of $C_a$ for all $a\ge 0$. Conversely  
the vanishing of $C_a$ for all $a\ge 0$ implies a differential (difference)
equation for the $\hat{C}_a$ which implies that $\hat{C}_a\propto \hat{C}_0
\equiv 0$. Thus the the sets of constraints are indeed equivalent. Next,
integrating (summing) by parts
\be \label{3.12}
\sum_{a\ge 0}\; f^a\; C_a=
\sum_i\; f^{\infty,i} \hat{C}_{\infty,i}-f^{0,i}\hat{C}_{0,i}
+\sum_{a\ge 0}\; \hat{f}^a\; \hat{C}_a,\; 
\hat{f}^a=\sum_{b\ge 0}\; \gamma^a_b\; f^b-D[\delta^{.}_b f^b]^a
\ee
where we have denoted a second boundary by the label $\infty$.
Imposing stability of $G^a=q^a-\tau^a=0,\;a>0$ 
under gauge transformations imposes $\hat{f}^a=0,\; a>0$ which is 
a {\it homogeneous, linear} 
differential (difference) equation for $f^a$ which can be solved in the form
$f^{r,i}=\lambda^j\;\tilde{f}_j^{r,i},\;r\ge 0$ where $\lambda^i$ are free 
parameters
while the invertible 
propagator $\tilde{f}$ is uniquely determined by that differential 
equation and 
say the initial condition $\tilde{f}_i^{0,j}=\delta_i^j$. 
We can now compute the reduced 
Hamiltonian $H$ that follows from these gauge fixing conditions acting on 
functions $F=F(\hat{R})$
\ba \label{3.13}
\{H,F\} &:=& \{\sum_{a\ge 0}\; f^a\; C_a,F\}_{C=G=f-\lambda\cdot \tilde{f}=0}
=\sum_i\; \{f^{\infty,i} \hat{C}_{\infty,i}-
f^{0,i}\hat{C}_{0,i},F\}_{C=G=f-\lambda\cdot \tilde{f}=0}
\nonumber\\
&=& 
\sum_i\;[f^{\infty,i} \{\hat{C}_{\infty,i},F\}
-f^{0,i}\{\hat{C}_{0,i},F\}]_{C=G=f-\lambda \tilde{f}=0}
=\sum_{i,j}\; \lambda^j\; (\tilde{f}_j^{\infty,i})_{G=C=0} 
\{p_{\infty,i}+\hat{h}_{\infty,i},F\}]_{C=G=0}
\nonumber\\
&=&\sum_{i,j}\; \lambda^j\; 
(\tilde{f}_j^{\infty,i})_{G=C=0}\; \{\hat{h}_{\infty,i},F\}]_{G=0}
=\sum_{i,j}\; \lambda^j\; (\tilde{f}_j^{\infty,i})_{G=C=0}\; 
\{(\hat{h}_{\infty,i})_{G=0},F\}
\ea
where in the first step we used that the Poisson bracket must act on
$\hat{C}$ in order not to vanish and that $\hat{f}=0$ when 
$f=\lambda\cdot\tilde{f}$ so that only the boundary term in (\ref{3.12}) 
survives,
in the second step again that the Poisson bracket must involve $\hat{C}$, 
in the third that $\hat{C}_{0,i}$ vanishes identically, in the fourth that 
$F$ does not depend on $q^a,\; a>0$ and in particular not on $q^{\infty,i}$
and in the fifth that as $F$ does not depend on $p_a,\; a>0$ we can impose 
$G^a=0,\; a>0$ before computing the Poisson bracket. 

Since (\ref{3.13}) is supposed to be of the form $\{H,F\}$ and since 
$\tilde{f}^\infty_{G=C=0}$ is a non-trivial matrix valued 
function on phase space we 
are forced to choose $\lambda$ to be of the form
\be \label{3.14}
\lambda^i:=(\frac{d\chi(z)}{dz_i})_{z=E}\;
([\tilde{f}^\infty_{G=C=0}]^{-1})_j^i;\;\;\; 
E_i(\hat{R}):=\hat{h}_{\infty,i}(\{q^b=\tau^b\}_{b>0},\hat{R}),\;
\ee
where $\chi$ is an arbitrary function which plays 
the same role as in the above version using the boundary term 
formulation while $E_i(\hat{R})$ is the analog of 
$j_A^\ast$. Note that $\lambda$ correctly does 
not depend on $q^a, a>0$. Then 
\be \label{3.15}
H(\hat{R})=\chi(E(\hat{R}))
\ee
is the physical Hamiltonian. 

Thus the time independent gauge fixing which 
is possible when the constraints depend on derivatives (differences) of the 
momenta is similar to the case of the time dependent gauge fixing when 
the constraints just depend algebraically on the momenta: The fact that 
derivatives are involved releases a several parameter freedom in the gauge 
fixed 
Lagrange multipliers $f^a$ parametrised by $\lambda$ while in the case of 
just algebraic dependence at least one such 
parameter (called $t$) must be supplied in the 
gauge fixing condition itself. The final physical Hamiltonian in both cases 
can be written in terms of the solution $p_a=-\hat{h}_a,\;y_j=-\hat{h}_j$ of
$C_a=C_j=0$ restricted to $x^j=0, q^a=\tau^a$. The difference is that 
in the time independent case $q^a=\tau^a$ only for $a>0$ corresponding to 
the fact that there is ``one constraint less'' per index $i$, i.e. that 
$p_{0,i}$ cannot be 
solved for due to the presence of the derivatives (differences). 
Another difference is that 
while in the case of time dependent gauge fixings the function $\chi$ in 
(\ref{3.15}) is linear in the $h_a$ and involves all $a$, while in the time 
independent case it involves only $a=\infty$ and is not necessarily linear.
The possible forms of $\chi$ will be determined by the condition that 
(\ref{3.14}) yields an allowed smearing function, i.e. with the specified 
decay behaviour such that $H$ is in fact well defined, i.e. both finite 
in value as functionally differentiable.\\ 
\\
Remarks:\\
1.\\
The discussion unveils the origin of the additional observables $p_0,q^0$
(e.g. mass, charge, angular momentum plus conjugate configuration variables
in case of black holes)
which are not present in the case without derivatives of momenta 
(e.g. in cosmology).\\
2.\\
The presence
of momentum derivatives offers the possibility to consider time independent
gauge fixings thus freeing the physical Hamiltonian from any explicit
time dependence, i.e. the reduced system is conservative.\\ 
3.\\
We note that in case that the constraints are not solvable algebraically for 
the momenta but rather involve a differential (difference) equation
we can compute the solution $S_a(\{q^b\}_{b>0},\hat{R})$ by the 
Picard-Lindel\"of method \cite{PL} i.e. by transforming the differential 
(difference) equation into an integral (sum) equation
\be \label{3.16}
p_{r,i}=p_{0,i}+I_{r,i}(p,q,R),\; I_{0,i}(p,q,R)=0
\ee
and iterating the right hand side. The perturbative scheme explained in the 
next section relies on the computation of a solution of (\ref{3.16}) 
at zeroth order with respect to $X,Y$ i.e. by solving it for $X=Y=0$. 
This does not introduce any non-polynomial dependence of the solution 
$p_a(0)$ so obtained on $X,Y$ but in general it will involve 
$\{q^a\}_{a\ge 0},p_0,Q^A,P_A$ non-polynomially which is also true in the case 
without momentum derivatives (differentials) (except that no dependence 
on $p_0$ is present). As the perturbative scheme of the next section 
only relies on the polynomial dependence of $C_a,C_j$ with respect to 
$x,y,X,Y$, it also applies in presence of momentum differentials. See below 
for cases where non-polynomial dependence at least of $Q,P$ can 
be avoided.\\
4.\\
As the physical Hamiltonian involves taking $r\to \infty$ in (\ref{3.16}) 
it typically involves an integral (sum) over all $r$ which therefore 
provides additional ``smearing'' of products of operator valued 
distributions upon quantisation of the fields $Q,P,X,Y$ 
involved and thus improves the chance that the physical 
Hamiltonian itself be promotable to an operator. For the same 
reason, in the time dependent case one will choose $\tau^a(t)\not=0$ for
all $a$ in order to have an integral (sum over $a$) involved. \\
5.\\
The perturbative algorithm of the next section directly computes $h_a,h_j$ 
and thus $E_i=\sum_{n=0} E_{i,(n)}$ in (\ref{3.14}) 
perturbatively and if $\chi$ in (\ref{3.15}) is linear in $E_i$,
this also directly computes $H$ 
perturbatively. If $\chi$ is non-linear, an additional perturbative 
Taylor expansion of $\chi$ is required, schematically 
\be \label{3.17}
H=\chi(E_{(0)})
+\sum_{k=0}\; \frac{1}{k!}\; \chi^{(k)}(E_{(0)})\;[\sum_{l=1} E_{(l)}]^k
=\chi(E_{(0)}+\chi^{(1)} \; E_{(1)}
+[\frac{1}{2}\;\chi^{(2)} \; E_{(1)}^2+\chi^{(1)} \; E_{(2)}]+...
\ee
which again constructs $H$ perturbatively.\\
6.\\
If the perturbative algorithm of the next section involves solving 
differential (difference) equations then this can be done using 
the Picard-Lindel\"of method by reformulating the problem as an 
integral equation as sketched above. This involves in addition 
an expansion of the integrand around 
the integration constant $p_0$ thereby constructing $\hat{h}_\infty$
as an iterated integral over polynomial expressions now not 
only in $X,Y$ but also in $Q,P$ with 
possibly non-polynomial dependence on $p_0,q^0$. This makes 
a quantisation of $H$ conceivable even if $Q,P,X,Y$ are fields rather 
than finitely many degrees of freedom. Moreover, since then the 
interaction terms of the symmetric background observables 
$Q,P$ with the non-symmetric perturbation observables $X,Y$ is such 
that $Q,P$ apear only in integrated form e.g. in a mass term for $X,Y$ 
there is a chance that the corresponding backreaction can be treated 
with the methods of space adiabatic perturbation theory even if 
$Q,P$ are fields \cite{SAPT,ST} with $Q,P$ playing the role of the 
``slow'' variables and $X,Y$ that of the ``fast''. The intuition 
behind this adiabatic split of the observables is that the $Q,P$ are 
by construction the {\it average} over the action of the symmetry group 
on all observables $Q,P.X,Y$ which make them similar to the centre of 
mass mode (mass weighted {\it average}) in classical mechanics 
with the much larger total mass as compared 
to the individual ones. The challenge is that 
in the field theory case the Weyl quantisation scheme for $Q,P$ on which 
SAPT relies has 
to be extended to infinitely many degrees of freedom (this is 
in contrast to cosmology where the phase space of the $Q,P$ is finite 
dimensional). A possible regularisation consists in working with 
a mode cut-off on the $Q,P$ degrees of freedom and thus to make 
the phase space of the $Q,P$ finite dimensional. Then at the end one 
removes that cut-off using methods of renormalisation.\\
7.\\
In both the time dependent and time independent cases we have considered 
gauge fixing conditions of the form $G^a=q^a-\tau^a,\; G^j=x^j-\rho^j$
where $\tau,\rho$ {\it are constant on the phase space}. More general 
gauge fixing conditions of the form $G=\tilde{G}-\tau$ are possible 
where $\tau$ is again constant on phase space but $\tilde{G}$ 
is a non-constant 
function such that the matrix $\Delta:=\{C,G\}$ is invertible. Then most of the 
statements of section (\ref{s3.1}) remain valid if one 
replaces $\hat{C}$ by $\Delta^{-1}\cdot C$ just that exact relations become 
weak (i.e. they hold modulo $C=0$). However, in this case it is 
difficult to make concrete statements about the form of the reduced 
Hamiltonian. Moreover, if $\tilde{G}$ also involves the momenta $p,y$, then
the perturbative scheme of the next section would break down because 
we could not disentangle the solution of the constraints in terms of $p,y$
at fixed $q,x$ from imposition of the gauge conditions. It is therefore
important to stick to such ``simple'' gauge conditions that involve only 
the configuration degrees of freedom.

\section{Perturbation theory in terms of Dirac observables}
\label{s4}

In principle, given the theory provided in the previous section, 
the strategy is clear: We declare $Q,P,X,Y$ as the true degrees of freedom
and $q,x;p,y$ as the pure gauge and constrained degrees of freedom 
respectively, compute the reduced Hamiltonian $H(t)=H(Q,P,X,Y;t)$ 
using gauge fixing conditions 
$G^a=\tau^a-q^a,\;G^j=\rho^j-x^j$ and, e.g. for explictly timed dependent 
gauge fixings, a one parameter curve 
$t\mapsto (\tau(t),\rho(t))$ therein and finally apply standard Hamiltonian 
perturbation theory to $H(t)$ with respect to the non-symmetric 
observables $X,Y$. This way the difficult question of how to perform 
perturbation theory on the unreduced phase space while keeping the first 
class property of the perturbed constraints is avoided altogether.
The question that is left open is whether this is 
feasible. In this section we develop the reduction viewpoint which 
is technically more convenient.\\
\\
We thus start again with the gauge system defined by  (\ref{2.8}), 
(\ref{2.10}), (\ref{2.11}). By (\ref{2.10}), the leading order contribution
to $C_j$ is $C_{j(1)}$ which linear in the perturbations $x,y,X,Y$ 
while the leading order contribution to $C_a$ is $C_{a(0)}$ which 
is independent of the perturbations. We 
assume that we can solve $C_j$ exactly for $y_j$ in the form (modulo 
global issues in phase space, see the previous section) 
\be \label{4.1}
C_j(p,q,x,y,P,Q,X,Y)=0 \Leftrightarrow\;\;
\tilde{C}_j=y_j+\tilde{h}_j(p,q,x;P,Q,X,Y)=0
\ee
We insert the solution $y_j=-\tilde{h}_j$ into $C_a$ and assume that we can 
solve it excactly for $p_a$ in the form 
\be \label{4.2}
(C_a)_{y=-\tilde{h}}=0 \;\;
\Leftrightarrow\;\; \hat{C}_a=p_a+h_a(q,x;P,Q,X,Y)
\ee
Then we insert $p_a=-\hat{h}_a$ into $\tilde{C}_j$ and find 
\be \label{4.3}
\hat{C}_j=y_j+h_j(q,x;P,Q,X,Y),\;
h_j=(\tilde{h}_j)_{p=-h}
\ee
By the theory provided in the previous section, the physical Hamiltonian 
is given by 
\be \label{4.4}
H(t;P,Q,X,Y)=
\dot{\tau}^a(t)\;h_a(\tau(t),\rho(t);P,Q,X,Y) 
+\dot{\rho}^j(t)\;h_j(\tau(t),\rho(t);P,Q,X,Y)
\ee
which can be Taylor expanded with respect to the observables $X,Y$ 
to any desired order provided we can construct $H(t;P,Q,X,Y)$ sufficiently 
explicitly.

Whether this is the case depends critically on the form of the constraints.
If $C_a,C_j$ depend non-polynomially on the momenta $p,y$ this will 
be impossible. Fortunately in physical applications this is not the 
case if the pair $(C_a,C_j)$ results from the symmetric -- non-symmetric 
split of constraints $C_\mu$ which depend polynomially on all 
momenta $i_{{\cal A}}$. This will be the case 
if the underlying Lagrangian depends on 
finitely many time derivatives by the Ostrogradski
method \cite{Ostrogradski}.
Since the symmetry split does not affect polynomiality, for these 
theories $C_a,C_j$ depend polynomially on all momenta and in particular
on $p,y$. 

While not necessary for what follows, we note that the constraints 
that we encounter in GR are not polynomial in the configuration 
coordinates. However, they can in fact also be made polynomial with respect
to all variables if we multiply them by phase space dependent factors 
that are classically non vanishing (certain powers of the determinant of the 
intrinsic metric). While this would make a significant difference in 
the Dirac approach to constrained systems in which one quantises the 
constraints on the unreduced phase space, this has {\it no influence}
on the reduced Hamiltonian in the reduced phase space approach. The reason 
for this is that given constraints $C_\mu$, a non-singular matrix 
$M_\mu\;^\nu$ and gauge fixing conditions $G^\mu$ such that 
$\Delta_\mu\;^\nu=\{C_\mu,G^\nu\}$ is non-singular, the smearing 
functions $f^\mu,\tilde{f}^\mu$ of $C_\mu,\tilde{C}_\mu:=M_\mu\;^\nu\; C_\nu$
are determined as $\hat{f},\hat{\tilde{f}}$
by fixing the gauge and respectively satisfy  
$\hat{f}^\nu\;\Delta_\nu\;^\mu=\dot{\tau}^\mu
=\hat{\tilde{f}}^\nu\;\tilde{\Delta}_\nu\;^\mu$ where 
$\tilde{\Delta}=\Delta\cdot M$. Then the reduced Hamiltonian is computed from
$\{H,F\}:=\{f^\mu\; C_\mu,F\}_{C=G=f-\hat{f}=0}$ and this coincides with
$\{\tilde{H},F\}:=\{\tilde{f}^\mu\; \tilde{C}_\mu,
F\}_{\tilde{C}=G=\tilde{f}-\hat{\tilde{f}}=0}$ thanks 
to invertibility of $M$. This has the following significance in perturbation
theory: In the polynomial form, the perturbative expansion of the constraints
is obviously a finite series while in the non-polynomial form that series 
is infinite with little control on the radius of convergence. Accordingly
in the polynomial form the task to compute the perturbative expansion 
is itself {\it possible to all orders}, it is exactly available and 
as such is a {\it non-perturbative} expression. One just chooses to write 
the constraints in variables that are adapted to the symmetry under 
discussion, {\it without dropping terms}. For instance, in vacuum GR it 
is possible to write the constraints as polynomials of order ten. In this form
one can carry out mode expansions and mode integrals discussed 
in section \ref{s2} in closed form, it just requires elementary methods from 
harmonic analysis on the symmetry group. Thus the shift of 
focus from the constraints to the reduced phase space leads to a 
{\it significant simplification and improvement} in the computational 
effort and the error control. Nevertheless, the perturbative 
expansion of the physical Hamiltonian involves an infinite series unless
all momenta appear only linearly. This is because for higher polynomial 
degree one needs to take (square) roots and their perturbative expansion 
yields an infinite series. The algorithm of theorem \ref{th4.1} displayed below 
computes that series directly perturbatively. Note that only in fortunate 
cases the non-perturbative 
expression of the polynomial form of the constraint may in fact allow to 
take those (square) roots exactly so that a non-perturbative expression 
for the reduced Hamiltonian is available, on the other hand, the presence 
of square roots may complicate its quantisation.       
 
Coming back to the task of solving the constraints 
for the momenta whether or not the constraints are polynomial also 
in the configuration variables, still the task of solving the system of 
polynomial equations (in the momenta)
$C_a=C_j=0$ for $p_a=-h_a,\; y_j=-h_j$ appears to be hopeless:
Solving systems of polynomials 
defining algebraic varieties
is the main task in the field of algebraic 
geometry \cite{AlgGeo} 
and already for a finite number of 
non-linear polynomials an extremely difficult task (unless 
the equations can be decoupled into polynomials in just one variable
of degree at most four) and an active field of 
research in pure mathematics. Since in our field theoretic setting we are 
dealing with an {\it infinite} number of polynomials, we are even leaving 
the terrain of known mathematics when trying to solve the infinite, coupled
system {\it exactly}. Even if we could do so, since typically 
an infinite number of non-linear equations are involved, there are an 
infinite number of sign choices to be made when selecting the various 
roots. What saves the day is that here we are interested in a {\it
perturbative setting} and we will show that in this case one can find 
a unique solution perturbatively. That solution also delivers the 
perturbation theory for the physical Hamiltonian in {\it one stroke}. 

Systems of polynomial equations are equivalent to larger systems of 
polynomial equations whose degree is at most two. For instance 
the cubic system $x^3+2x^2+3y=0, xy^2+4xy+5x=0$ in two variables 
and two equations is equivalent to the quadratic system 
$u=xy,v=x^2,xv+2v+3y=0, yu+4u+5x=0$ in four variables and four equations. 
Thus we may assume that $C_a,C_j$ 
depend at most quadratically on the momenta $p,y,P,Y$. In GR this is even 
the case without enlarging the system. Accordingly we isolate the 
dependence of $C_a,C_j$ on $p,y$ using the following notation:
\ba \label{4.5}
C_a &=& U_a+K_a^b \; p_b+L_a^j\;y_j
+A_a^{bc}\; p_b\; p_c
+B_a^{jk}\; y_j\; y_k
+C_a^{bj}\; p_b\; y_j
\nonumber\\
C_j &=& V_j+M_j^a \; p_a+N_j^k\;y_k
+D_j^{ab}\; p_a\; p_b
+E_j^{kl}\; y_k\; y_l
+F_j^{ak}\; p_a\; y_k
\ea
where
\ba \label{4.6}
U_a &=& u_a
+u_a^A\; P_A+u_a^J\; Y_J
+u_a^{AB}\; P_A\;P_B+u_a^{JK}\; Y_J\; Y_K+u_a^{AJ} P_A\; Y_J
\nonumber\\
V_j &=& v_j
+v_j^A\; P_A+v_j^J\; Y_J
+v_j^{AB}\; P_A\;P_B+v_j^{JK}\; Y_J\; Y_K+v_j^{AJ} P_A\; Y_J
\nonumber\\
K_a^b &=& k_a^b+k_a^{bA}\; P_A+k_a^{bJ} Y_J
\nonumber\\
L_a^j &=& l_a^j+l_a^{jA}\; P_A+l_a^{jJ} Y_J
\nonumber\\
M_j^a &=& m_j^a+m_j^{aA}\; P_A+m_j^{aJ} Y_J
\nonumber\\
N_j^k &=& n_j^k+n_j^{kA}\; P_A+n_j^{kJ} Y_J
\ea
The notation is as follows: The coefficients $U,V$ and $K,L,M,N$ 
and $A,B,C,D,E,F$ respectively displayed in (\ref{4.5}) 
are second order and first order and 
zeroth order polynomials in $P_A,Y_J$ respectively as displayed  
in  (\ref{4.6}) where the various coefficients coefficients $u,v,k,l,m,n$
(which are functions of $q,x,Q,X$ only) 
are known in terms of their Taylor expansion in terms of 
$x,X$ using standard perturbation theory on the unreduced phase space.
 Without loss of generality $A_a^{bc},D_j^{bc}$ and 
$B_a^{kl},\;E_j^{kl}$ respectively
are symmetric in $b,c$ and $k,l$ respectively. 

As before, we will denote by $()_{(n)}$ the n-th order monomial 
in the Taylor expansion of $()$ wrt $x,y,X,Y$ with coefficient functions that 
depend only on $q,Q$. Thus e.g.
\be \label{4.7}
U_{a(n)} = u_{a(n)}
+u_{a(n)}^A\; P_A+u_{a(n-1)}^J\; Y_J
+u_{a(n)}^{AB}\; P_A\;P_B+u_{a(n-2)}^{JK}\; Y_J\; Y_K
+u_{a(n-1)}^{AJ} \; u_a^{AJ} P_A\; Y_J
\ee
The motivated identities $C_{a(1)}=C_{j(0)}=0$ (\ref{2.10}) now translate into 
\ba \label{4.8}
C_{a(1)} &=& U_{a(1)}+K_{a(1)}^b \; p_b+L_{a(0)}^j\;y_j
+A_{a(1)}^{bc}\; p_b\; p_c
+C_{a(0)}^{bj}\; p_b\; y_j=0
\nonumber\\
C_{j(0)} &=& V_{j(0)}+M_{j(0)}^a \; p_a
+D_{j(0)}^{ab}\; p_a\; p_b=0
\ea
for all $p,y$. Taking zeroth, first and second derivatives at $p=y=0$ yields
\be \label{4.9}
U_{a(1)}=K_{a(1)}^b=L_{a(0)}^j=A_{a(1)}^{bc}=C_{a(0)}^{bj}=0,\;\;
V_{j(0)}=M_{j(0)}^a=D_{j(0)}^{ab}=0
\ee
To solve $C_a=C_j$ perturbatively for $p_a=-h_a,\;y_j=-h_j$ we expand 
\be \label{4.10}
p_a=\sum_{n=0}^\infty\; p_a(n),\; 
y_j=\sum_{n=1}^\infty\; y_j(n)
\ee
where $p_a(n):=-h_{a(n)},\;y_j(n):=-h_{j(n)}$ are n-th order monomials wrt 
$x,X,Y$ with coeffients depending on $q,Q,P$ which are to be determined. 
Note that $p_a(0)\not=0=y_j(0)$ consistent with the perturbative scheme that 
requires $p_a, y_j$ respectively to be a zeroth and first order quantity
respectively.
\begin{Theorem} \label{th4.1} ~\\
Suppose that a solution $p_a(0)$ of $C_{a(0)}=0$ can be found and that the 
$x,X,Y$ independent matrices
\be \label{4.11}
R_a^b(q,Q,P):=K_{a(0)}^b+2\;A_{a(0)}^{cb}\; p_c(0),\;\;\; 
S_j^k(q,Q,P):=N_{j(0)}^k+F_{j(0)}^{ck}\; p_c(0) 
\ee
are non-degenerate. Then $C_a=0=C_j$ has a unique (up to the choice of 
root $p_a(0)$) perturbative solution. In particular $p_a(1)=0$. 
\end{Theorem}
Remarks:\\
i.\\
Since $C_a(0)=0$ is one of the {\it exact} field equations of the {\it purely 
symmetric} system, the assumption that a solution $p_a(0)$ exists is 
physically well justified because imposing the symmetry was 
{\it motivated} by the desire to find exact solutions. \\
ii.\\
The assumed regularity of $R,S$ is a selection criterion for the split 
of the full canonical pair 
$(k,i)$ into $(q,p),(x,y),(Q,P),(X,Y)$. Note that the regularity of 
$R,S$ is a condition that one also imposes in usual first and second 
order perturbation theory and since $R,S$ do not receive corrections at higher
oder, their inversion which enters into the iteration as displayed 
below does not get more involved at higher orders.\\ 
iii.\\
The solution constructed below is formal in the sense that nothing is known 
about the radius of convergence of the corresponding power series 
in $x,X,Y$ (pointwise in $q,x,Q,P,X,Y$). Such a convergence analysis is beyond 
the scope of the present manuscript and is reserved for future analysis.\\
iv.\\
The proof displays the advantage of working with constraints polynomial
in all variables when available as the range of the sums in (\ref{4.13}) 
gets automatically truncated in terms of the polynomial degree 
of the coefficient functions.\\
\\ 
\begin{proof}:\\
The proof is by induction over 
$N=0,1,2,...$ in solving $C_a^{(N)}=C_j^{(N)}=0$. We do this by inserting 
the expansions (\ref{4.10}) into the decomposition (\ref{4.5}) and isolating 
homogenous perturbation orders. We start with the $N=0,1$ 
cases and use (\ref{4.9})
\ba \label{4.12}
C_{a(0)} &=& U_{a(0)}+K_{a(0)}^b \; p_b(0)
+A_{a(0)}^{bc}\; p_b(0)\; p_c(0)=0
\nonumber\\
C_{a(1)} &=& 
U_{a(1)}+
K_{a(1)}^b \; p_b(0)+
K_{a(0)}^b \; p_b(1)+
L_{a(0)}^j\;y_j(1)
+A_{a(1)}^{bc}\; p_b(0)\; p_c(0)
+2A_{a(0)}^{bc}\; p_b(0)\; p_c(1)
\nonumber\\
&& +C_{a(0)}^{bj}\; p_b(0)\; y_j(1)
\nonumber\\
&=& [K_{a(0)}^b +2A_{a(0)}^{cb}\; p_c(0)]\; p_b(1)
=R_a^b \; p_b(1) \equiv 0
\nonumber\\
C_{j(0)} &=& V_{j(0)}+M_{j(0)}^a \; p_a(0)
+D_{j(0)}^{ab}\; p_a(0)\; p_b(0)
\equiv 0
\nonumber\\
C_{j(1)} &=& 
V_{j(1)}+
M_{j(1)}^a \; p_a(0)
+M_{j(0)}^a \; p_a(1)
+N_{j(0)}^k\;y_k(1)
+D_{j(1)}^{ab}\; p_a(0)\; p_b(0)
+2D_{j(0)}^{ab}\; p_a(0)\; p_b(1)
\nonumber\\
&& +F_{j(0)}^{ak}\; p_a(0)\; y_k(1)
\nonumber\\
&=& [V_{j(1)}+M_{j(1)}^a \; p_a(0)
+D_{j(1)}^{ab}\; p_a(0)\; p_b(0)]
+[N_{j(0)}^k+F_{j(0)}^{ak}\; p_a(0)]\; y_k(1)
\nonumber\\
&=& [V_{j(1)}+M_{j(1)}^a \; p_a(0)
+D_{j(1)}^{ab}\; p_a(0)\; p_b(0)]
+S_j^k\; y_k(1)=0
\ea
The first equation is supposed to be solved by $p_a(0)$ which thus is a
known function depending only on $q,Q,P$. Thus also the matrices 
$R,S$ in (\ref{4.11}) only depend on $q,Q,P$. The second equation is 
equivalent to $p_a(1)=0$ by virtue of the assumed regularity of $R$. The 
third equation is already identically satisfied due to (\ref{4.9}) and 
the fourth equation uniquely determines $y_j(1)$ as a homogeneous 
linear function of $x,X,Y$ by virtue of the assumed regularity 
of $S$. 

Thus $p_a(0),\;p_a(1)=0,\;y_j(0)=0,\;y_j(1)$ are all determined by the 
the equations $C_{a(0)}=C_{a(1)}=C_{j(0)}=C_{j(1)}=0$. The idea is 
that when inluding the next terms $C_{a(2)}, C_{j(2)}$ to 
obtain $C_a^{(2)}, C_j^{(2)}$, then solving $C_a^{(2)}=C_j^{(2)}=0$
we can account for that by adding corrections $p_a(2),y_j(2)$ to 
$p_a(0),y_j(1)$. This can be repeated. We thus assume 
inductively that for some $N\ge 2$ all $p_a(n),y_j(n)$ for $n=0,..,N-1$
have been found iteratively 
by solving $C_{a(n)}=C_{j(n)}=0$ for $n=0,..,N-1$. We now 
isolate $C_{a(N)}$ and $C_{j(N)}$ exploiting $p_a(0)=y_j(1)=0$ and 
the relations (\ref{4.9}). We display the excluded values 
of the integers $n,r,s\ge 0$ where we sum over the occurring values 
$n,r$ or $n,r,s$ respectively subject to $n+r=N$ or
$n+r+s=N$ respectively 
\ba \label{4.13}
C_{a(N)} &=& U_{a(N)}
+\sum_{n\not=1, r\not=1,n+r=N} \; K_{a(n)}^b \; p_b(r)
+\sum_{n\not=0, r\not=0,n+r=N} \; L_{a(n)}^j \;y_j(r)
\nonumber\\
&&
+\sum_{n\not=1, r,s\not=1,n+r+s=N} \;  A_{a(n)}^{bc}\; p_b(r)\; p_c(s)
+\sum_{r,s\not=0,n+r+s=N} \;B_{a(n)}^{jk}\; y_j(r)\; y_k(s)
\nonumber\\
&&
+\sum_{n\not=0, r\not=1,s\not=0,n+r+s=N} \;  C_{a(n)}^{bj}\; p_b(r)\; y_j(s)
\nonumber\\
C_{j(N)} &=& V_{j(N)}
+\sum_{n\not=0, r\not=1,n+r=N} \; M_{j(n)}^a \; p_a(r)
+\sum_{r\not=0,n+r=N} \; N_{j(n)}^k\;y_k(r)
\nonumber\\
&& +\sum_{n\not=0, r,s\not=1,n+r+s=N} \; D_{j(n)}^{ab}\; p_a(r)\; p_b(s)
+\sum_{r,s\not=0,n+r+s=N} \;E_{j(n)}^{kl}\; y_k(r)\; y_l(s)
\nonumber\\
&& +\sum_{r\not=1,s\not=0,n+r+s=N} \; F_{j(n)}^{ak}\; p_a(r)\; y_k(s)
\ea
Due to $n+r=N$ or $n+r+s=N$ and $n\ge 0$ the top order of $r$ or $s$ in 
(\ref{4.13}) can be at most $N$ and only if $n=0$ is an allowed value.
If both $r,s$ cannot take the value $0$ then also neither of $r,s$ can take 
the value $N$ and if one of $r,s$ cannot take the value $0$ the other 
cannot take the value $N$. Thus, 
in $C_{a(N)}$ the only terms that allow for $r=N$ 
or $s=N$ are the second and fourth 
while 
in $C_{j(N)}$ the only terms that allow for $r=N$ 
or $s=N$ are the third and sixth. Isolating those and denoting the 
remainder as $C'_{a(N)},\;C'_{j(N)}$ respectively which involves only 
the already known values of $p_a(n),y_j(n),\;n\le N-1$ we find 
\ba \label{4.14}
0 &=& 
C'_{a(N)}+[K_{a(0)}^b+2 A_{a(0)}^{cb} \; p_c(0)]\;p_b(N)  
=C'_{a(N)}+R_a^b\;p_b(N)  
\nonumber\\
0 &=& C'_{j(N)}+[N_{j(0)}^k+F_{j(0)}^{ck} p_c(0)]\; y_k(N) 
= C'_{j(N)}+S_j^k\; y_k(N) 
\ea
which can be solved uniquely for $p_a(N),\;y_j(N)$ due to the 
regularity of both $R,S$.
\end{proof}
Note:\\
i. \\
It seems that we have introduced more degrees of freedom $p_a(n),y_j(n);\;
n=0,1,2,...$ than we had originally (i.e. just $p_a,y_j$) and that 
we solved more equations $C_{a(n)}=0, C_{b(n)}=0$ than we had originally
(i.e. only $C_a=0, C_j=0$).
However this is not the case: The $p_a(n),y_j(n)$ are just 
auxiliary constructs, of interest is only their sum. When summed 
up in (\ref{4.10}) up to order $N$ yielding $-h_a^{(N)}, -h_j^{(N)}$ 
respectively, by virtue of the above construction, they
give zero up to a term of order $N+1$ when inserted into $C_a, C_j$.
This works because for each $N$ the equations $C_{a(N)}=0=C_{j(N)}$ 
involve only the variables $p_a(n),y_j(n);\; n\le N$ and thus are not 
affected by adding higher order corrections to $h_a^{(n)},h_j^{(n)}$.  
Thus we immediately get the $N-th$ order approximation of the reduced 
Hamiltonian
\be \label{4.14a}
H^{(N)}:=\dot{\tau}^a [h_a^{(N)}]_{q=\tau,x=\rho} 
+\dot{\rho}^j [h_j^{(N)}]_{q=\tau,x=\rho} 
\ee 
ii.\\
As detailed by the proof, the contributions $p_a(n),y_j(n)$ involve 
$n-th$ powers of the inverses of $R,S$ which are functions 
of $q,x,Q,P$. Since in the final Hamiltonian we are only interested 
in the evaluation at the prescribed values $q=\tau(t),x=\rho(t)$ 
installing the gauge fixing conditions, we thus find that the 
physical Hamiltonian depends on inverse powers of $R,S$ and thus 
inverse powers of functions of 
of the observables $Q,P$ while it depends only on positive powers 
of the observables $X,Y$. Negative powers in the physical 
Hamiltonian of just one of $Q,P$ are not problematic in quantum theory
as they just correspond to singular potentials and the corresponding 
operator can still be densely defined on a suitable domain. However,
negative powers of both $P,Q$ could be potentially problematic unless 
the functions of both $P,Q$ of which inverse powers occur have suitable 
properties. For example these functions could be  
bounded from below by a positive constant. This is indeed the case in some 
examples of interest, including cosmology. Obviously, nothing can be said 
in general about this issue, it requires a case by case analysis. 
A possibility consists in performing an additional power expansion in at least 
one of $Q,P$ around the integration constants $p_0$ of the previous 
section when they are present. \\    
iii.\\
The gauge condition $\rho^j=0$ is particularly simple and adopted by many 
practitioners, it is also used in the partial reduction approach described 
in the next section. In this case, the lowest order for which the 
backreaction Hamiltonian (\ref{4.14}) depends non-trivially on the 
gauge invariant perturbations $X,Y$ is $N=2$. To compute it, we need 
$p_a(0)=-h_{a(0)},p_a(1)\equiv 0,p_a(2)=-h_{a(2)},y_j(0)\equiv 0, y_j(1)$ 
while $y_j(2)$ is not needed to compute 
$H^{(2)}=\dot{\tau}^a\;[h_{a(0)}+h_{a(2)}]$. These can be computed iteratively
using theorem \ref{th4.1}. Using it, one finds the explicit formula
for $y_j(1)$ is displayed in (\ref{6.12}) assuming that $p_a(0)$ has been 
found. Thus, having $p_a(0),y_j(1)$ at our disposal we find from 
(\ref{4.13}), (\ref{4.14})
\be \label{4.15}
h_{a(2)}=(R^{-1})_a^b\;
[
U_{a(2)}+K_{a(2)}^b \; p_b(0)+ L_{a(1)}^j \;y_j(1)
+A_{a(2)}^{bc}\; p_b(0)\; p_c(0)
+B_{a(0)}^{jk}\; y_j(1)\; y_k(1)
+C_{a(1)}^{bj}\; p_b(0)\; y_j(1)
]
\ee

\section{Reduction in stages}
\label{s5}

In the first subsection we outline the theory for partial classical and 
partial quantum reduction for general gauge systems combined with 
perturbation theory. Analogous to section \ref{s3} we develop both the 
gauge invariance and the reduction viewpoint. The gauge invariance 
viewpoint will be used in the next subsection in order to compare
with \cite{Gomar} which uses the gauge invariance viewpoint applied 
to second order cosmological perturbation theory with backreaction.
We will show that the results of \cite{Gomar} are precisely 
embedded into the approach developed here, opening the avenue for 
generalisation to higher order. Hence,  
the reduction viewpoint will be used in section 
\ref{s6} in order to display the details of {\it third order} 
perturbation theory in this partially reduced context for a general 
gauge system (not only cosmology).

We show that there is generically an obstacle whenever 
there are more than one  unreduced remaining constraints when 
the latter are to be quantised which is one of the motivations 
to perform full reduction as compared to partial reduction. If on the 
other hand there is only one remaining constraint, then  
our perturbation theory applied in the partial reduction context, 
does not suffer from that 
quantisation obstacle, to arbitrary order. 
This is precisely the situation in cosmological 
perturbation theory.

It should be stressed that what is called a gauge invariant observable
in the context of this section should better be called {\it partially}
gauge invariant observable: They are just invariant with respect to the 
subset of constraints with respect to which the partial reduction is carried
out. This is in contrast to the previous section where observable means 
a fully gauge invariant object. This abuse of notation
common in cosmology arises for historical reasons: If one neglects 
backreaction then the unreduced constraint (there is only one in this case)
becomes a physical Hamiltonian and no longer has the status of a constraint,
in that sense the partially reduced variables are full observables.

\subsection{Partial reduction and perturbation theory}
\label{s5.1}
 
It maybe desirable to carry out the reduction only partially with respect 
to a non-trivial subset of the constraints. For example for practical 
reasons one may wish 
to carry out a classical reduction with respect to the chosen subset 
of constraints and a quantum reduction with respect to the remaining 
subset of constraints. In order that this works, the chosen subset 
must form a closed subalgebra in the constraint algebra. In what follows 
we display the theory for the case that the classical reduction be performed
with respect to the ``non-symmetric constraints'' $C_j$. This will 
enable us to compare a similar procedure developed in \cite{Gomar} for 
second order cosmological perturbation theory.\\
\\
We first note that the subalgebra condition is not automatically satisfied 
if the structure functions $\kappa_{jk}\;^a$ displayed in (\ref{2.11}) are 
non-trivial. Thus, the first step must be to pass to an equivalent set 
of constraints $\tilde{C}_a,\tilde{C}_j$ for which the corresponding 
$\tilde{\kappa}_{jk}\;^a$ all vanish. Given the theory layed out in section
\ref{s3}, a good candidate for this are the constraints 
$\tilde{C}_a=\hat{C}_a=p_a+h_a,\;\tilde{C}_j=\hat{C}_j=y_j+h_j$ 
which are even Abelian with respect to all of $\hat{C}_a,\hat{C}_j$. 
Since we want to keep the $C_a$ intact as much as possible
for the purposes of quantisation,
we may work instead with the equivalent constraint set 
$C_a,\; \tilde{C}_j:=\hat{C}_j$ where 
$\hat{C}_j=y_j+h_j(q,x,Q,P,X,Y)$ is obtained 
by solving $C_a$ for $p_a=-\tilde{h}_a(q,x,y,Q,P,X,Y)$ and then solving 
$C_j(q,p=-\tilde{h}(q,x,y,Q,P,X,Y),x,y,Q,P,X,Y)=0$ for 
$y_j=-h_j(q,x,Q,P,X,Y)$. However, as 
$\tilde{C}_j=\lambda_j^a \; C_a+\nu_j^k C_k$ 
and $C_j=(\nu^{-1})_j^k[\tilde{C}_k-\lambda_k^a C_a]$
for certain functions $\lambda$ and invertible $\nu$ 
and given (\ref{2.11}) we now find 
\ba \label{5.1}
\{C_a,C_b\} &=& 
(\kappa')_{ab}\;^c \; C_c+(\kappa')_{ab}\;^j\; \tilde{C}_j
\nonumber\\
\{C_a,\tilde{C}_j\} &=& 
(\kappa')_{aj}\;^b \; C_b+(\kappa')_{aj}\;^k\; \tilde{C}_k
\nonumber\\
\{\tilde{C}_j,\tilde{C}_k\} &=& 0 
\ea
for certain new structure functions $\kappa'$ which can 
be explicitly computed from $\kappa,\lambda,\nu$.

\subsubsection{Gauge invariance viewpoint}
\label{s5.1.1}

As we wish to get rid of the constraints $\tilde{C}_j$ classically
we want to still modify the constraint $C_a$ into $\tilde{C}_a$ 
in order to achieve for the resulting yet further 
modified structure functions that
$\tilde{\kappa}_{ab}\;^j=\tilde{\kappa}_{aj}\;^b=
\tilde{\kappa}_{aj}\;^k=0$. Given the theory layed out in section (\ref{s3})
this may be achieved as follows: Since the $\tilde{C}_j$ form a closed,
Abelian subalgebra we may apply (\ref{3.3}) with the partial set 
of Abelian Hamiltonian vector 
fields $X_j=\{\tilde{C}_j,.\}$ instead of the full 
Abelianised set $X_\mu$. That is, for any function $F$ on phase space
\be \label{5.2}
O_F:=F+\sum_{n=1}^\infty\; 
\frac{1}{n!}\; \tilde{G}^{j_1}..\tilde{G}^{j_n}\;
[X_{j_1}\cdot ..X_{j_n}\cdot F]
\ee
with the gauge fixing condition $\tilde{G}^j:=-x^j$ (i.e. we set 
$\tau=0$ in (\ref{3.3})).

Applied to $F=C_a$ we note that $X_j\cdot C_a$ is a linear combination 
of constraints via the second relation in (\ref{5.1}), thus also 
$X_j\cdot X_k \cdot C_a$ is a linear combination of constraints etc. 
It follows that 
\be \label{5.3}
\tilde{C}_a:=O_{C_a}
\ee
together with $\tilde{C}_j$ forms an equaivalent set of constraints
as the zeroth order term with respect to the $x^j$ in (\ref{5.3}) 
starts with $C_a$.  Moreover,
(\ref{5.3}) 
enjoys the following properties (see the list of properties displayed in 
section (\ref{s3}))
\ba \label{5.4}
\{\tilde{C}_a,\tilde{C}_b\} &=& O_{\{C_a,C_b\}^\ast}
\nonumber\\
\{\tilde{C}_j,\tilde{C}_a\} &=& 0
\ea
The last relation says that $\tilde{C}_a$ is an observable with respect 
to the $\tilde{C}_j$ by construction of the ``projector'' $F\mapsto O_F$.
The first relation can be further worked out using the Dirac bracket
defined by the pair $(\tilde{C}_j, \tilde{G}^j)$
\be \label{5.5}
\{F,F'\}^\ast=\{F,F'\}-
[\{F,\tilde{C}_j\}\; \{\tilde{G}^j,F'\}
-\{F',\tilde{C}_j\}\; \{\tilde{G}^j,F\}]
\ee
For $F=C_a, F'=C_b$ both correction terms in (\ref{5.5}) is a linear 
combination of constraints $C_a, \tilde{C}_j$. It follows that 
also $\{\tilde{C}_a,\tilde{C}_b\}$ is a linear combination of constraints 
$\tilde{C}_a,\tilde{C}_j$. The constraints $\tilde{C}_a$ can also 
be written (see section \ref{s3})
\be \label{5.6} 
\tilde{C}_a=C_a(q\to O_q, p\to O_p, x\to 0, y\to \tilde{C}-O_h,
Q\to O_Q, P\to O_P, X\to O_X, Y\to O_Y)
\ee
Since by construction $\{O_F,O_{F'}\}=\{O_F,O_{F'}\}^\ast$ and since with 
respect to the Dirac bracket we may set $\tilde{C}_j=0$ before or after 
evaluating the bracket, we may pass to the partial constraint surface 
$\tilde{C}_j=0$ and henceforth forget about the degrees of freedom 
$x^j,\tilde{C}_j$ which form a canonical pair and the partially reduced 
phase space is coordinatised by $O_F$ with $F\in \{q,p,Q,P,X,Y\}$ while 
we set $\tilde{C}_j$ strongly to zero in (\ref{5.6}), that is
\ba \label{5.7} 
\tilde{C}_a &=& C_a(q\to O_q, p\to O_p, x\to 0, y\to -O_h,
Q\to O_Q, P\to O_P, X\to O_X, Y\to O_Y),\;
\nonumber\\
O_{h_j} &=& h_j(q\to O_q,x\to 0,Q\to O_Q, P\to O_P, X\to O_X, Y\to O_Y)
\ea
Moreover, since $x^j$ has vanishing Poisson brackets with 
$F\in \{q,p,Q,P,X,Y\}$ it follows that 
\be \label{5.8}
\{O_F,O_{F'}\}=
O_{\{F,F'\}^\ast}=O_{\{F,F'\}}=\{F,F'\}
\ee
for $F,F'\in \{q,p,Q,P,X,Y\}$ since then $\{F,F'\}=$const. so that the 
$O_F,O_{F'}$ remain conjugate variables. Thus Poisson algebraically 
nothing is changed under the substitution $F\mapsto O_F$ for functions 
of $q,p,Q,P,X,Y$ only, i.e. on this sector 
of the phase space the map $O$ is an exact  Poisson isomorphism or canonical 
transformation.

\subsubsection{Reduction viewpoint}
\label{s5.1.2}

Alternatively to this gauge invariance (with respect to the 
$\tilde{C}_j$) viewpoint we may 
consider the partially reduced phase space where the reduction 
is with respect to the $\tilde{C}_j$. This is completely analogous 
to the previous section and the summary is that 
we can forget about the degrees of freedom
$x,y$ and the constraints $C_j$ altogether and just keep the 
constraints 
$C_a$ as long as we modify them into 
\be \label{5.8a}
\tilde{C}_a(q,p,Q,P,X,Y)=
C_a(q,p,x=0,y=-h(q,x=0,Q,P,X,Y),Q,P,X,Y)
\ee 
Here $h_j$ was constructed 
perturbatively in the previous section. 
In particular these modified constraints 
close among 
themselves 
\be \label{5.9}
\{\tilde{C}_a,\tilde{C}_b\}=\tilde{\kappa}_{ab}\;^c\; \tilde{C}_c
\ee
for certain $\tilde{\kappa}$ which are explicitly computed
in the next section, see (\ref{6.2}).
To have a classical first class algebra (\ref{5.9}) 
is a prerequisite for a consistent quantisation. 
The problem is that only the 
{\it exact} $\tilde{C}_a$ are granted to close among themselves, 
i.e. only when 
we invoke the entire series (\ref{4.10}) that defines $h_j$ and not only
its $N-$th order truncation. The only exception is the 
case that the index $a$ takes only one value. If that is not the case and 
if we construct $\tilde{C}_a$ perturbatively, then the closure of 
the approximated constraints will in general be
violated thus prohibiting a consistent quantisation 
which, however, was the whole motivation for the only partial reduction
with respect to $C_j$. Thus, the perturbation theory for $h_j$
and thus $\tilde{C}_a$ developed  
in the previous section is granted to be be viable for the partial 
reduction approach developed in this section only for systems for 
which there is only one 
symmetric constraint $C_a$ as is the case for instance in cosmology but 
not for black holes. Thus, whenever the index range of $a$ comprises at
least two values we must resort to the full reduction process developed 
in section \ref{s4}. 

If on the other hand, if there is only one symmetric 
constraint $C_a$ then the $N-$th order 
truncation\\ 
$C_a^{(N)}(q,p,Q,P,X,Y)$ of the function  
\be \label{5.10}
C_a(q,p,x=0,y=-h^{(N)}(q,x=0,Q,P,X,Y),Q,P,X,Y)
\ee
with respect to $X,Y$, where $h^{(N)}_j$ is the series (\ref{4.10}) 
truncated at order $N$, 
will be a well motivated starting point for quantisation. The $N-th$ order 
truncation of (\ref{5.10}) is readily computed since 
$C_a(p,q,x,y,P,Q,X,Y)$ is polynomial in $y$ as displayed in 
(\ref{4.5}). Thus, e.g. for $N\ge 3$ using either the previous 
or the present section one would get a viable theory describing cosmological 
non-Gaussianities resulting from the self-interactions of the non-symmetric 
(that is in this case, inhomogeneous) degrees of freedom $X,Y$ while
there are only a finite number of degrees of freedom in the list 
$q,p,Q,P$ (the homogeneous modes of all matter and geometry degrees of 
freedom). We display the details in section \ref{s6} for $N=3$.

\subsection{Comparison with the literature at second order}
\label{s5.2}

We first review the quite elaborate procedure developed in \cite{Gomar}
and then show that it embeds very naturally into the context of the 
previous subsection, thus explaining the symplecto-geometric origin of the 
procedure followed in \cite{Gomar}. We only consider 2nd order perturbation
theory in this section and keep the index $a$ for easier comparison 
with the rest of this paper but keep in mind that eventually we are 
only interested in univalent index range of $a$.\\
\\
We start again with the constraints $C_a,C_j$ and, as motivated in section 
(\ref{s2}), truncate them at second and first order respectively 
\be \label{5.11}
C_{a(1)}=0, \; C_{j(0)}=0, C^{(2)}_a=C_{a(0)}+C_{a(2)},\;
C_j^{(1)}=C_{j(1)}
\ee
We also recall the symmetric and non-symmetric brackets (\ref{2.12}) as well 
as the following identies derived in section \ref{s2} 
\ba \label{5.12}
&& \{C_{a(0)},C_{b(0)}\}_S=  
\{C_{a(0)},C_{b(0)}\}=\kappa_{ab(0)}\;^c\; C_{c(0)},\;\;
\nonumber\\
&& \{C_{j(1)},C_{k(1)}\}_{\bar{S}}=  
\kappa_{jk(0)}\;^a\; C_{a(0)}
\nonumber\\
&&\{C_{j(1)},C_{a(0)}\}_S+\{C_{j(1)},C_{a(2)}\}_{\bar{S}}
=\kappa_{ja(0)}\;^k C_{k(1)}+\kappa_{ja(1)}\;^b C_{b(0)}
\ea
Since the $C_{j(1)}$ do not close among each other, 
the first step of \cite{Gomar} is to select some first 
order canonical variable $L^j$ 
subject to the conditions 
\be \label{5.13}
\{C_{j(1)},L^k\}_{\bar{S}}=\delta_j^k,\;\{L^j,L^k\}_{\bar{S}}=0
\ee
In the present situation, a possible choice is 
\be \label{5.13a}
L^j:=x^k\; (\sigma^{-1})_k^j,\;\sigma_j^k:=\{C_{j(1)},x^k\}_{\bar{S}}
\ee
Whenever such functions can be found we can consider the ``improved'' 
constraints 
\be \label{5.14}
C'_{j(1)}=C_{j(1)}+\frac{1}{2}\kappa_{jk(0)}\; L^k\; C_{a(0)}
\ee
which enjoy the property (note that all zero order quantities have vanishing 
$\{.,.\}_{\bar{S}}$ bracket) 
\be \label{5.15}
\{C'_{j(1)},C'_{k(1)}\}_{\bar{S}}= 
\{C_{j(1)},C_{k(1)}\}_{\bar{S}}
+\frac{1}{2}[\kappa_{km(0)}\;^a\; \{C_{j(1)},L^m\}_{\bar{S}} 
-\kappa_{jm(0)}\;^a\; \{C_{k(1)},L^m\}_{\bar{S}}]\; C_{a(0)}=0
\ee
i.e. they are Abelian albeit only with respect to the non-symmetric 
bracket. We may replace $C_{j(1)}$ by $C'_{j(1)}$ free of charge in 2nd 
order perturbation theory because the totally constrained Hamiltonian 
reads 
\be \label{5.16}
f^a C_a^{(2)}+g^j C_{j(1)}
=[f^a-\frac{1}{2} g^j L^k\kappa_{jk(0)}\;^a] C_{a(0)}
+ f^a C_{a(2)}+g^j C'_{j(1)}
=[f^a-\frac{1}{2} g^j L^k\kappa_{jk(0)}\;^a] C_a^{(2)}
+g^j C'_{j(1)}+O(3)
\ee
where the third order correction can be dropped in 2nd order perturbation 
theory. In other words, the substitution of $C_{j(1)}$ by (\ref{5.14}) 
can be induced by a redefinition of $f^a$ up to a higher order correction
of the Hamiltonian.

If we are interested in backreaction, it is not enough that (\ref{5.15}) 
holds with respect to the non-symmetric bracket only, it should 
hold with respect to the full bracket. Therfore the second step of 
\cite{Gomar} consists in constructing a canonical transformation 
on the full phase space, at least to second order perturbation 
theory, from 
the canonical coordinates $(q,p,x,y,Q,P,X,Y)$ to new canonical coordinates 
$(q',p',x',y',Q',P',X',Y')$ with $y'_j:=C'_{j(1)},\; (x')^j:=L^j$.
In other words the $\{.,.\}_{\bar{S}}$ 
canonical pair $(L^j,C'_{j(1)})$ as displayed in 
(\ref{5.13}), (\ref{5.13a}), (\ref{5.14}) and (\ref{5.15}) is completed 
to a canonical 
pair on the full phase space which triggers corresponding changes 
in the other canonical coordinates. To derive this transformation 
one notices that \cite{Gomar}
\be \label{5.17} 
(X')^J:=X^J-L^j\{C_{j(1)},X^J\}_{\bar{S}}
=X^J-L^j\{C'_{j(1)},X^J\}_{\bar{S}},\;\;
Y'_J:=Y_J-L^j\{C_{j(1)},Y_J\}_{\bar{S}}
=Y_J-L^j\{C'_{j(1)},Y_J\}_{\bar{S}},\;\;
\ee
are such that $(x',y',X',Y')$ have canonical $\{.,.\}_{\bar{S}}$ brackets 
among each other since $\kappa_{jk(0)}\;^a \; C_{a(0)}$ and $\sigma_j^k$ 
have vanishing $\{.,.\}_{\bar{S}}$ bracket. We will therefore use these 
$x',y',X',Y'$ as candidates and try to compute $q',p',Q',P'$ at least up 
to second order. 
To do this we compute the inversions of 
(\ref{5.13a}), (\ref{5.14}) and (\ref{5.17})
i.e. 
\ba \label{5.19}
&& x^j=\sigma_k^j\; (x')^k= (x')^k \{C'_{k(1)},x^j\}_{\bar{S}},\;\;
X^J=(X')^J+(x')^j\{C'_{j(1)},X^J\}_{\bar{S}}
\nonumber\\
&& Y_J=Y'_J+(x')^j\{C'_{j(1)},Y_J\}_{\bar{S}},\;\;
y_j=(\sigma^{-1})_j^k\;[
y'_k-(\sigma_{jk} \;x^k+\sigma_{jK}\; X^K+\sigma_j^K\; Y_K)]
\ea
where we have used that $C'_{j(1)}$ is linear in $x,y,X,Y$ with 
with coefficient functions $\sigma$ depending on $q,p,Q,P$
\be \label{5.20}
C'_{j(1)}=
\sigma_{jk} \;x^k+\sigma_j^k y^k+\sigma_{jK}\; X^K+\sigma_j^K\; Y_K
\ee
We will condense the notation by writing the variables in two groups 
corresponding to symmetric and non-symmetric degrees of freedom
\be \label{5.21}
r^\alpha=(q^a,Q^A),
s_\alpha=(p_a,P_A),
u^\rho=(x^j,X^J),
v_\rho=(y_j,Y_J)
\ee
and may write the content of (\ref{5.19}) compactly as 
\be \label{5.22}
u^\rho=M^\rho_\lambda\; (u')^\lambda+M^{\rho\lambda}\; v'_\lambda,\;
v_\rho=_{\rho\lambda}\; (u')^\lambda+M_\rho^\lambda\;  v'_\lambda
\ee
where the matrices $M$ depend only on $r,s$ and define 
by construction a linear canonical transformation 
$(u',v')\mapsto (u,v)$ at constant $(r,s)$. We plug (\ref{5.22}) 
into the symp[lectic potential
\be \label{5.18}
2\Theta= s_\alpha \; dr^\alpha-ds_\alpha \; r^\alpha
+v_\rho\; du^\rho- dv_\rho\; u^\rho
\ee
where the exterior derivative $d$ acts on all phase space variables.
If we collect the terms in which $d$ acts on $u',v'$ we know that 
this just replaces the $(u,v)$ in (\ref{5.18}) by $(u',v')$ by virtue 
of $M$ being canonical for each $r,s$. Thus 
\ba \label{5.23}
2\Theta &=& s_\alpha \; dr^\alpha-ds_\alpha \; r^\alpha
+v'_\rho\; d(u')^\rho- dv'_\rho\; (u')^\rho
+2\;(\Delta s_\alpha)\; dr^\alpha-2\; ds_\alpha \; (\Delta r^\alpha)
\nonumber\\
&=& 
(s_\alpha+(\Delta s)_\alpha)\; d(r^\alpha+(\Delta r)^\alpha)
-d(s_\alpha+(\Delta s)_\alpha)\; (r^\alpha+(\Delta r)^\alpha)
+v'_\rho\; d(u')^\rho- dv'_\rho\; (u')^\rho
+d(.)+O(4)
\nonumber\\
&& 2(\Delta s)_\alpha=v_\rho\; u^\rho_{,r^\alpha}-v_{\rho,r^\alpha}\; u^\rho,
\;\; 
-2(\Delta r)^\alpha= v_\rho\; u^\rho_{,s_\alpha}-v_{\rho,s_\alpha}\; u^\rho
\ea
where in the last step we dropped a total differential and a term of 
fourth order in $u,v$ because $\Delta r,\Delta s$ are of second order.
The notation in the last line means that $u,v$ are considered as functions 
of the independent coordinates $u',v',r,s$. 
Thus up to second order 
\be \label{5.24}
(r')^\alpha=r^\alpha+(\Delta r)^\alpha,\;
s'_\alpha=s_\alpha+(\Delta s)_\alpha,\;
\ee
where in the expressions for $\Delta r, \Delta s$ we reexpress $u',v'$ in 
terms of $u,v$. A closer look at $\Delta r, \Delta s$ shows that 
\be \label{5.25}
2(\Delta r)^\alpha=\{v_\rho,r^\alpha\}_S u^\rho-v_\rho\; 
\{u^\rho, r^\alpha\}_S,\;\;
2(\Delta s)_\alpha=\{v_\rho,s_\alpha\}_S u^\rho-v_\rho\; 
\{u^\rho, s_\alpha\}_S
\ee
where again $u,v$ are functions of $u',v',r,s$ as displayed in 
(\ref{5.22}). We once more compactify the notation and write 
$t^{{\cal A}}=(r^\alpha,s_\alpha),\; w^{{\cal M}}=(u^\rho,v_\rho)$ 
and introduce the antisymmetric, constant 
symplectic structure matrices $\Omega^S,\Omega^{\bar{S}}$ defined 
by 
\be \label{5.26}
2\Theta=
\Omega^S_{{\cal A}{\cal B}} t^{{\cal A}}\; dt^{{\cal B}}
+\Omega^{\bar{S}}_{{\cal M}{\cal N}} w^{{\cal M}}\; dw^{{\cal N}}
\ee
with inverse $\Omega_S^{{\cal A}{\cal B}}$ i.e. 
$\Omega_S^{{\cal A}{\cal C}}\;(\Omega_S)_{{\cal C}{\cal B}}=
\delta^{{\cal A}}_{{\cal B}}$ and similar for 
$\Omega_{\bar{S}}$. It is standard but also not difficult to check 
directly that with the non-vanishing 
Poisson bracket convention 
$\{y_j,x^k\}_{\bar{S}}=\delta_j^k,\;  
\{Y_J,X^K\}_{\bar{S}}=\delta_J^K\;
\{p_a,q^b\}_S=\delta_a^b,\;  
\{P_A,Q^B\}_S=\delta_A^B$ we have   
\be \label{5.27}
\{F,G\}_S=-\Omega_S^{{\cal A}{\cal B}}\; F_{,{\cal A}}\; G_{,{\cal B}},\;\;
\{F,G\}_{\bar{S}}=-\Omega_{\bar{S}}^{{\cal M}{\cal N}}\; 
F_{,{\cal M}}\; G_{,{\cal N}}
\ee
with $F_{,{\cal A}}:=\partial F/\partial t^{{\cal A}},\;
F_{,{\cal M}}:=\partial F/\partial w^{{\cal M}}$. 
In this notation (\ref{5.22}) becomes 
\be \label{5.28}
w^{{\cal M}}=M^{{\cal M}}_{{\cal N}}\; (w')^{{\cal N}}
\ee
and 
(\ref{5.25}) becomes ($w$ is considered a function 
of $w',t$) 
\ba \label{5.28a}
-2(\Delta t)^{{\cal A}}
&=& 
\Omega^{\bar{S}}_{{\cal M}{\cal N}}\; w^{{\cal M}}\; 
\{w^{{\cal N}},t^{{\cal A}}\}_S
\nonumber\\ 
&=& 
\Omega^{\bar{S}}_{{\cal M}{\cal N}}\; w^{{\cal M}}\; 
\{M^{{\cal N}}_{{\cal P}},t^{{\cal A}}\}_S\; (w')^{\cal P}
\nonumber\\ 
&=&
\Omega^{\bar{S}}_{{\cal M}{\cal N}}\; w^{{\cal M}}\; 
\{M^{{\cal N}}_{{\cal P}},t^{{\cal A}}\}_S\; 
(M^{-1})^{{\cal P}}_{{\cal Q}}\; w^{\cal Q}
\ea
which expresses the corrections $\Delta t$ in terms of $w,t$ as $M$ 
is also a function of $t$. Next 
\ba \label{5.28b}
-2(\Delta t)^{{\cal A}}
&=& -\Omega^{\bar{S}}_{{\cal M}{\cal N}}\; w^{{\cal M}}\; 
M^{{\cal N}}_{{\cal P}}
\{(M^{-1})^{{\cal P}}_{{\cal Q}},t^{{\cal A}}\}_S\; 
w^{\cal Q}
\nonumber\\
&=& -\Omega^{\bar{S}}_{{\cal M}{\cal N}}\; w^{{\cal M}}\; 
M^{{\cal N}}_{{\cal P}}
\{(w')^{{\cal P}},t^{{\cal A}}\}_S\; 
\ea
where in the second step we used that $\{.,.\}_S$ does not act on $w$
and reintroduced $w'=M^{-1} w$ considered as a function of $w,t$. Since $M$ 
is canonical wrt $\Omega^{\bar{S}}$ we have 
\be \label{5.29}
\Omega^{{\bar S}}_{{\cal M}{\cal N}} \;
M^{{\cal M}}_{{\cal P}}
M^{{\cal N}}_{{\cal Q}}=\Omega^{{\bar S}}_{{\cal P}{\cal Q}}
\;\; \Leftrightarrow \;\;
\Omega^{{\bar S}}_{{\cal M}{\cal N}} \;
M^{{\cal N}}_{{\cal Q}}=
\Omega^{{\bar S}}_{{\cal P}{\cal Q}}\; (M^{-1})^{{\cal P}}_{{\cal M}}
\ee
and thus we can cast (\ref{5.28b}) into the simpler form 
\be \label{5.30}
-2(\Delta t)^{{\cal A}}
=-
\Omega^{{\bar S}}_{{\cal Q}{\cal P}}\; (M^{-1})^{{\cal Q}}_{{\cal M}}
\; w^{{\cal M}}\; 
\{(w')^{{\cal P}},t^{{\cal A}}\}_S\; 
=- (w')^{{\cal M}}\; \Omega^{{\bar S}}_{{\cal M}{\cal N}}\; 
\{(w')^{{\cal N}},t^{{\cal A}}\}_S\; 
\ee
Reintroducing $x',y',X',Y'$ we get 
\be \label{5.31}
-2(\Delta t)^{{\cal A}}
=-y'_j \{(x')^j,t^{{\cal A}}\}_S
+(x')^j \{y'_j,t^{{\cal A}}\}_S
-Y'_J \{(X')^J,t^{{\cal A}}\}_S
+(X')^J \{Y'_J,t^{{\cal A}}\}_S
\ee
To summarise we have shown that 
\ba \label{5.32}
y'_j &:&=C'_{j(1)}=C_{j(1)}+\kappa_{jk(0)}\;^a\; C_a(0)\; L^k/2
\\
(x')^j &:=& L^j=(\sigma^{-1})_k\;^j \; x^k,\;\; \sigma_j^k=
\{C_{j(1)},x^k\}_{\bar{S}}
\nonumber\\
(X')^J &:=& X^J- (x')^j \{C'_{j(1)},X^J\}_{\bar{S}},\;
Y'_J := Y_J- (x')^j \{C'_{j(1)},Y_J\}_{\bar{S}},\;
\nonumber\\
F'&=& F+\frac{1}{2}[
y'_j \{(x')^j,f\}_S-(x')_j \{y'_j,f\}_S + 
Y'_J \{(X')^J,f\}_S-(X')^J \{Y'_J,f\}_S],\;F\in \{q^a, p_a, Q^A, P_A\}
\nonumber
\ea
defines a canonical transformation with respect to the full Poisson 
bracket, up to 2nd order in $x,y,X,Y$, i.e. the failure from being an 
exact canonical transformation is an at least third order correction.

Therefore $q',p',Q',P',X',Y'$ in (\ref{5.32}) define observables with 
respect to $C'_{j(1)}$ up to second order corrections. 
It is however not clear that $C_a^{(2)}=C_{a(0)}+C_{a(2)}$ in 
(\ref{5.11}) can be written in terms of these 2nd order observables 
up to higher order corrections. For this to be the case it would
be necessary that $\{C'_{j(1)},C_a^{(2)}\}$ is of second order modulo 
$C'_{k(1)},C^{(2)}_a$ terms. Given (\ref{5.12})
we compute the modifications of (\ref{5.12}) that result from 
the substitution $C_{j(1)}\to C'_{j(1)}$. Using 
$\hat{\kappa}_{jk(0)}\;^a=\frac{1}{2}\hat{\kappa}_{jl(0)}\;^a\; 
(\sigma^{-1})_k^l$ we find for the full brackets (note that $C'_{j(1)}$
is still homogeneous linear in $x,y,X,Y$)
\ba \label{5.33}
&&\{C'_{j(1)},C'_{k(1)}\}
= \{C'_{j(1)},C'_{j(1)}\}_{\bar{S}}+O(2)=O(2)
\nonumber\\
&& \{C'_{j(1)},C_a^{(2)}\}  
= \{C_{j(1)},C_a^{(2)}\}
+\{\hat{\kappa}_{jk(0}\;^b\;x^k\; C_{b(0)},C_a^{(2)}\}
\nonumber\\
&=& \{C_{j(1)},C_{a(0)}\}_S+\{C_{j(1)},C_{a(2)}\}_{\bar{S}}+O(3)
+\{\hat{\kappa}_{jk(0}\;^b\;x^k\; C_{b(0)},C_{a(0)}\}_S
+\{\hat{\kappa}_{jk(0}\;^b\;x^k\; C_{b(0)},C_{a(2)}\}_{\bar{S}}+O(3)
\nonumber\\
&=& \kappa_{ja(0)}\;^k C_{k(1)}+\kappa_{ja(1)}\;^b C_{b(0)}
\nonumber\\
&&
+x^k\;[
\{\hat{\kappa}_{jk(0}\;^b,C_{a(0)}\}_S \; C_{b(0)}
+\hat{\kappa}_{jk(0}\;^b\; \kappa_{ab(0)}\;^c C_{c(0)}]
+\hat{\kappa}_{jk(0}\;^b\;\{x^k,C_{a(2)}\}_{\bar{S}}\; C_{b(0)}+O(3)
\nonumber\\
&=:& 
\kappa_{ja(0)}\;^k\; C'_{k(1)}+
\tilde{\kappa}_{ja(1)}^b\; C_{b(0)}+O(3)
\ea
The $C'_{j(1)}$ close among themselves up to O(2) corrections by construction
but $C_a^{(2)}$ is not an observable with respect to $C'_{j(1)}$ up to 
O(2) corrections, there is an O(1) obstruction term proportional to 
$C_{a(0)}$. We thus modify $C_a^{(2)}$ by 
\be \label{5.34}
C_a^{\prime(2)}:=C_a^{(2)}+\lambda_{a(2)}^b C_{b(0)}
\ee
where the O(2) correction can be absorbed into a redefinition of the 
smearing function $(f')^a=f^a+\lambda_{b(2)}^a f^b$ up to higher order 
corrections since 
$(f')^a C_a^{(2)}-f^a C_a^{\prime(2)}=O(4)$. Then 
\ba \label{5.35}
&& \{C'_{j(1)},C_a^{\prime(2)}\}
=\kappa_{ja(0)}\;^k\; C'_{k(1)}+
\tilde{\kappa}_{ja(1)}^b\; C_{b(0)}+O(3)
+\{C'_{j(1)},\rho_{a(2)}^b C_{b(0)}\}  
\nonumber\\
&=& \kappa_{ja(0)}\;^k\; C'_{k(1)}+
[\tilde{\kappa}_{ja(1)}^b
+\{C'_{j(1)},\rho_{a(2)}^b\}_{\bar{S}}] C_{b(0)} +O(3)
\ea
The term proportional to $C_{b(0)}$ can be brought to vanish  
if the PDE system 
\be \label{5.36}
\tilde{\kappa}_{ja(1)}^b+\{C'_{j(1)},\rho_{a(2)}^b\}_{\bar{S}}=0
\ee
has a solution for which it is necessary that the 
corresponding integrability conditions 
\be \label{5.37}
2 \{C'_{[j(1)},
\tilde{\kappa}_{k]a(1)}^b+\{C'_{k](1)},\rho_{a(2)}^b\}_{\bar{S}}\}_{\bar{S}}=0
\ee
hold which, using the Jacobi identity with respect to $\{.,.\}_{\bar{S}}$
translates into 
\be \label{5.38}
2 \{C'_{[j(1)},
\tilde{\kappa}_{k]a(1)}^b\}_{\bar{S}}=O(2)
\ee
which is not granted to hold, not even, if the index range of $a,b$ is
uni-valent. We also note for completeness that 
\ba \label{5.39}
&&\{C_a^{(2)},C_b^{(2)}\}=
\{C_{a(0)},C_{b(0)}\}_S+
\{C_{a(0)},C_{b(2)}\}_S+
\{C_{a(2)},C_{b(0)}\}_S+
\{C_{a(2)},C_{b(2)}\}_{\bar{S}}+O(4)
\nonumber\\
& = &\kappa_{ab(0)}^c\;C_{b(0)}+
\kappa_{ab(0)}^c C_{b(2)}+
\kappa_{ab(2)}^c C_{b(0)}+
\kappa_{ab(1)}^j C_{j(1)}+
\kappa_{ab(0)}^j C_{j(2)}+
\nonumber\\
&=& \kappa_{ab(0)}\;^c\; C_c^{(2)}+O(2)
\ea
closes with itself but only up to O(2) corrections which are of the 
same order as $C_a^{(2)}$ which is discomforting. However, 
if we wish $C_a^{(2)}$ to be gauge invariant with respect to 
$C'_{j(1)}$ up to O(2) we must perform the substitution (if it exists) 
$C_a^{(2)}\to C_a^{\prime(2)}=\lambda_{a(2)}^b C_b(0)+C_a^{(2)}$ and 
then find 
\be \label{5.40}
\{C_a^{\prime(2)},C_b^{\prime(2)}\}
= \kappa_{ab(0)}\;^c\; C_c^{\prime(2)}+O(2)
\ee
where the additional O(2) corrections are all proportional to $C_{a(0)}$
with coefficients of the form 
$\{\lambda_{(2)},\lambda_{(2)}\}_{\bar{S}}$,
$\{\lambda_{(2)},C_{(2)}\}_{\bar{S}}, \lambda\kappa_{(0)}$. Of course,
(\ref{5.40}) vanishes when $a,b$ take only uni-valent range.\\
\\
Given these difficulties the procedure followed in the literature is to 
expand $C_a^{(2)}$ in terms of $q',p',x',y',Q',P',X',Y',P'$, to drop 
higher order terms and all terms proportional to $y'$ (corresponding 
to a redefinition of $g^j$) and all powers of $x'$. Dropping the 
$x'$ dependent terms is usually justified by arguing that $x'$ is pure gauge 
and thus these terms cannot be observable. The so modified $C'_{a(2)}$ 
then satisfies $\{C'_{j(1)},C^{\prime(2)}_a\}=O(2)$ by construction and 
trivially $\{C^{\prime(2)}_a,C^{\prime(2)}_b\}=0$ for uni-valent range of $a,b$.
However, the dropping of the $x'$ dependent terms lacks a more profound 
justification other than providing a consistent model. \\
\\
\\
Given the theory 
developed in section \ref{s3} and specialised to the current situation
in the previous subsection, one may wonder whether there is any connection
between (\ref{5.32}) and the observable ``projector'' $O_F$ in 
\ref{5.2}. This is far from obvious: \\
1. In (\ref{5.32}) 
an elaborate mixture and symmetric and non-symmetric brackets appears 
while in (\ref{5.2}) just involves the full Poisson bracket. \\
2. In (\ref{5.32}) the correction terms involve Poisson brackets 
of $F\in \{q,p,Q,P,X,Y\}$ with both $C'_{j(1)}$ and $(X')^J,Y'_J$ while 
in (\ref{5.2}) only Poisson brackets between $F$ and  
$\tilde{C}_j=y_j+h_j(q,x,Q,P,X,Y)$ are involved. \\
3. While both $C'_{j(1)}=C'_{j(1)}(q,p,x,y,QP,X,Y)$ 
and $\tilde{C}_j$ are related to the same original 
unperturbed $C_j$, they do not even depend on the same set of variables
($\tilde{C}_j$ does not depend on $p$). 

Remarkably, still a precise relation between (\ref{5.32}) and (\ref{5.2})
can be established, thereby unveiling the symplecto-geometric origin 
of the quite elaborate set of formulae (\ref{5.32}) and 
showing how to extend the framework of \cite{Gomar} to higher orders.
It also justifies  
why the constrained Hamiltionian $f^a C_a^{(2)}+g^j C_{j(1)}$ can be 
replaced by an equivalent one  which can be written just in terms of 
$q',p',y',Q',P',X',Y'$ while an explicit dependence on $x'$ drops out:
As we showed in the previous section, before engaging in perturbation 
theory, one can replace $C_a,C_j$ by equivalent 
constraints (thereby redefining $f^a, g^j$ in the Hamiltonian)
$\tilde{C}_a,\tilde{C}_j$ such that 
with respect to the full bracket
the $\tilde{C}_a$ close among themselves, the $\tilde{C}_a,\tilde{C}_j$ 
commute and the $\tilde{C}_j,\tilde{C}_k$ are Abelian. The exact reduction 
of the theory with respect to $\tilde{C}_j$ should therefore result 
in $\tilde{C}_a$ which is fully gauge invariant with respect to 
$\tilde{C}_j$ and thus cannot depend on $x^j$ (which is such that 
$\{\tilde{C}_j,x^k\}=\delta_j^k$) explicitly. Indeed this is precisely 
achieved in (\ref{5.6}) (partially gauge invariant viewpoint) or 
(\ref{5.8a}) (partial reduction viewpoint). As $x,x'$ are linearly related
in the perturbative treatment above, all explicit dependence 
on $x'$ in $C_a^{(2)}$ should be eliminated. 

A first hint that (\ref{5.32}) and (\ref{5.2}) cannot be unrelated is 
due to the appearance of the factor $1/2$ in $q'-q,p'-p,Q'-Q,P'-P$ 
which is missing in $X'-X,Y'-Y$: Such a factor would appear precisely 
in the second order term of the Taylor expansion in (\ref{5.2}). Also 
the expansion parameter in (\ref{5.2}) is $x^j$ while in (\ref{5.32}) 
for at least some of the terms it is $(x')^j$ and $x,x'$ 
are linearly related. To understand the factor $1/2$ from the point of view
of  (\ref{5.2}) let us temporarily assume (to be justified below) that 
we may sustitute $(x')_j,y'_j=C'_{j(1)}$ by $x^j, \tilde{C}_{j(1)}$. Let 
$X_j=\{\tilde{C}_{j(1)},.\}$ then 
\be \label{5.41}
O_F=F-x^j \; \{\tilde{C}_{j(1)},F\} 
+\frac{1}{2} x^j\; x^k\; \; \{\tilde{C}_{j(1)},\tilde{C}_{k(1)},F\}\}
+O(3)
\ee
where we dropped higher order terms for second order perturbation theory: 
Since $x$ is of first oder, thus no matter what the result of the $n-$th
order Poisson bracket calculation is, the terms proportional to 
$x^n$ in (\ref{5.2}) are at least of order $n$. We determine the terms 
at most of order two in (\ref{5.41}) for a first order function $F_{(1)}$,
i.e. a linear function of $x,y,X,Y$ with coefficients depending on
$q,p,Q,P$ and a zeroth order function $F_{(0)}$ respectively. 
We have  
$\{\tilde{C}_{j(1)},F_{(1)}\}=\{\tilde{C}_{j(1)},F_{(1)}\}_{\bar{S}}
+O(2)$ and thus 
the second term in (\ref{5.41}) is contributing 
a quadratic correction. On the other hand    
\be \label{5.42}
\{\tilde{C}_{j(1)},\{\tilde{C}_{k(1)},F_{(1)}\}\}  
=\{\tilde{C}_{j(1)},O(0)+O(2)\}  
=\{\tilde{C}_{j(1)},O(0)\}_S+\{\tilde{C}_{j(1)},O(2)\}_{\bar{S}}+O(3)
=O(1)  
\ee
thus the third term in (\ref{5.41}) is of third order and can be dropped.
Next 
$\{\tilde{C}_{j(1)},F_{(0)}\}=\{\tilde{C}_{j(1)},F_{(0)}\}_S=O(1)$ 
and thus the second term in (\ref{5.41}) is a quadratic correction and thus 
\be \label{5.42a}
\{\tilde{C}_{j(1)},\{\tilde{C}_{k(1)},F_{(0)}\}\}  
=\{\tilde{C}_{j(1)},O(1)\}  
=\{\tilde{C}_{j(1)},O(1)\}_{\bar{S}}+O(2)
\ee
Therefore the third term which contains the factor 
$1/2$ in (\ref{5.41}) is also quadratic and does 
contribute when $F$ is of zeroth order! 

To see 
explicitly why the third term in (\ref{5.41}) is mandatory 
in order that $O_F$ be a second order observable when $F=F_{(0)}$ we compute 
assuming $\{\tilde{C}_{j(1)},x^k\}=\delta_j^k$ (justified below) 
\be \label{5.43}
\{\tilde{C}_{j(1)},F-x^k \{\tilde{C}_{k(1)},F\}\}
=-x^k\{\tilde{C}_{j(1)},\{\tilde{C}_{k(1)},F\}\}
\ee
which is O(2) for $F=F_{(1)}$ but only O(1) for $F=F_{(0)}$. Thus 
for $F=F_{(0)}$ we include the third term in (\ref{5.41}) and use
$\{\tilde{C}_j,\tilde{C}_k\}=O(2)$ (justified below). Then
\ba \label{5.44a}
&& \{\tilde{C}_{j(1)},O_F\}
=\{\tilde{C}_{j(1)},F\}-\{\tilde{C}_{j(1)},x^k \; \{\tilde{C}_{k(1)},F\}\}
+\frac{1}{2} 
\{\tilde{C}_{j(1)},x^k\; x^l\; \; \{\tilde{C}_{k(1)},\{\tilde{C}_{l(1)},F\}\}\}
\nonumber\\
&=&
-x^k\;\{\tilde{C}_{j(1)},\{\tilde{C}_{k(1)},F\}\}
+\frac{1}{2}[
x^k \; \{\tilde{C}_{k(1)},\{\tilde{C}_{j(1)},F\}\}\}
+x^l \; \{\tilde{C}_{j(1)},\{\tilde{C}_{l(1)},F\}\}\}
\nonumber\\
&& +x^k x^l \{\tilde{C}_{j(1)},\{\tilde{C}_{k(1)},\{\tilde{C}_{l(1)},F\}\}\}]
\nonumber\\
&=&
\frac{1}{2}\; x^k\;[
\{\tilde{C}_{k(1)},\{\tilde{C}_{j(1)},F\}\}\}
-\{\tilde{C}_j,\{\tilde{C}_{k(1)},F\}\}]+O(3)
\nonumber\\
&=&
\frac{1}{2}\; x^k\;
\{\tilde{C}_{k(1)},\{\tilde{C}_{j(1)}\},F\}\}
+O(3)
=O(3)
\ea
where we used the Jacobi identity.\\
\\
Summarising, to second order 
\be \label{5.44}
O_{X^J}=X^J-x^j \{\tilde{C}_{j(1)},X^J\},\;
O_{Y_J}=Y_J-x^j \{\tilde{C}_{j(1)},Y_J\}
\ee
already look very close to $(X')^J,Y'_J$ in (\ref{5.32}) while 
for $F\in \{q^a,p_a,Q^A,P_A\}$ the quantities $O_F$ and $F'$ in (\ref{5.32})
are not obviously related except that they involve a factor $1/2$.
To see how (\ref{5.2}) and (\ref{5.32}) are related nevertheless we must 
carry out the programme of the previous subsection to second order. 
The first step is to calculate $\tilde{C}_j=y_j+h_j(q,x,Q,P,X,Y)$
to first order where $h_j$ is computed perturbatively in section \ref{s4}.
Thus we suppose to have found a solution $p_a(0)$ (a function
of $q,Q,P$ of $C_a(0)=0)$ and 
must solve $C_j=0$ to first order in $y$ which is explicitly given by 
(see (\ref{4.12}) where what is denoted by $S_j^k$ there is denoted by 
$\sigma_j^k$ here)
\be \label{5.45}
\tilde{C}_{j(1)}=y_j+\{(\sigma^{-1})_j^k[\sigma_{jk}\; x^k
+\sigma_{jJ} \; X^J+  
+\sigma_j^J \; Y_J]\}_{p=p(0)}=\{(\sigma^{-1})_j^k\; [C'_{j(1)}]\}_{p=p(0)}
\ee  
where the notation (\ref{5.20}) was used. Note that 
$C'_{j(1)}=C_{j(1)}$ when $p=p(0)$ since these two functions differ by a 
term proportional to $C_{a(0)}$ which vanishes at $p=p(0)$ by definition.

This provides the first connection between the formalisms. The second is that
$(x')^j=\sigma_k^j x^k$ so that 
\be \label{5.46}
[(x')^j \; C'_{j(1)}]_{p=p(0)}=x^j \tilde{C}_{j(1)}
\ee
It follows that to second order
\ba \label{5.47}
&& O_{X^J}=X^J- x^j\;\{\tilde{C}_{j(1)},X^J\}
=X^J- x^j\;\{\tilde{C}_{j(1)},X^J\}_{\bar{S}}
=X^J- (x')^j\;\{[C'_{j(1)}]_{p=p(0)},X^J\}_{\bar{S}}
\nonumber\\
&=& X^J- (x')^j\;\{[C'_{j(1)}]_{p=p(0)},X^J\}_{\bar{S}}
=(X')^J_{p=p(0)}
\ea
and similarly $O_{Y_J}=[Y'_J]_{p=p(0)}$. 

Recall $y'_j=C'_{j(1)}$ and let 
$\tilde{y}_j=(\sigma^{-1})_j^k y'_k$ so that 
$\tilde{C}_{j(1)}=[\tilde{y}_j]_{p=p(0)}$. For $F\in \{q,p,Q,P\}$ 
note that all
brackets with $F$  are automatically $\{.,.\}_S$ brackets
which do not act on $x,y,X,Y$. Then 
\ba \label{5.47a}
&& -2(F'-F)=
y'_j \;\{(x')^j,F\}-(x')^j \;\{y'_j,F\}
+ Y'_J\;\{(X')^J,F\}-(X')^J\;\{Y'_J,F\}
\\
&=&
y'_j \;\{(x')^j,F\}-(x')^j \;\{y'_j,F\}
- Y'_J\;\{(x')^j \{y'_j,X^J\}_{\bar{S}},F\}
+(X')^J\;\{(x')^j\{y'_j,Y_J\}_{\bar{S}},F\}
\nonumber\\
&=&
y'_j \;\{(x')^j,F\}- x^k[
(\sigma^{-1})_k^j\{y'_j,F\}
+Y'_J\;\{(\sigma^{-1})_k^j \{y'_j,X^J\}_{\bar{S}},F\}
-(X')^J\;\{(\sigma^{-1})_k^j\{y'_j,Y_J\}_{\bar{S}},F\}
]
\nonumber\\
&=&
y'_j \;\{(x')^j,F\}- x^k[
\{\tilde{y}_k,F\}+
(\sigma^{-1})_k^j\; \tilde{y}_l\{\sigma_j^l,F\}
+Y'_J\;\{\{\tilde{y}_k,X^J\}_{\bar{S}},F\}
-(X')^J\;\{\{\tilde{y}_k,Y_J\}_{\bar{S}},F\}
]
\nonumber\\
&=&
y'_j \;\{(x')^j,F\}- x^k[
\{\tilde{y}_k,F\}
-y'_j\{\sigma^{-1})_k^j,F\}
+Y'_J\;\{\{\tilde{y}_k,X^J\}_{\bar{S}},F\}
-(X')^J\;\{\{\tilde{y}_k,Y_J\}_{\bar{S}},F\}
]
\nonumber\\
&=& 2 y'_j \;\{(x')^j,F\}
- x^k
[\{\tilde{y}_k,F\}
+Y'_J\;\{\{\tilde{y}_k,X^J\}_{\bar{S}},F\}
-(X')^J\;\{\{\tilde{y}_k,Y_J\}_{\bar{S}},F\}
]
\nonumber\\
&=& 2 y'_j \;\{(x')^j,F\}
\nonumber\\
&& - x^k
[
\{A_{kj},F\}\; x^j+\{B_{jJ},F\}\; X^J+\{C_j^J,F\}\; Y_J
+(Y_J+x^j\; B_{j})\;\{C_k^J,F\}
+(X^J-x^j\; C_j^J)\;\{B_{kJ},F\}
]
\nonumber\\
&=& 2 y'_j \;\{(x')^j,F\}
- x^k
[
\{A_{kj},F\}\; x^j+2\{B_{jJ},F\}\; X^J+2\{C_j^J,F\}\; Y_J
+x^j\;[B_{j})\;\{C_k^J,F\}-C_j^J\;\{B_{kJ},F\}]
]
\nonumber
\ea
where in the last step we used the abbreviations
\be \label{5.48}
\tilde{y}_j=y_j+(\sigma^{-1})_j^k[\sigma_{kl}\; x^l
+\sigma_{kJ}\; X^J
+\sigma_k^J\; Y_J)=:y_j+A_{jk}\; x^k+B_{jJ}\; X^J+C_j^J\; Y_J
\ee
On the other hand with $\bar{A}_{jk}:=[A_{jk}]_{p=p(0)}$ and similar 
for $\bar{B}_{jJ}, \bar{C}_j^J$ as well as 
$\bar{y}_j:=[\tilde{y}_j]_{p=p(0)}=\tilde{C}_{j(1)}$ we have 
\ba \label{5.49}
&& -2(O_F-F)
=2x^k \{\bar{y}_k,F\}-x^k\; x^j\;
\{\bar{y}_k,\{\bar{y}_j,F\}\}
\nonumber\\
&=& 2x^k \{\bar{y}_k,F\}-x^k\; x^j\;
\{\bar{y}_k,\{\bar{y}_j,F\}\}_{\bar{S}}+O(4)
\nonumber\\
&=& 
2x^k \{\bar{y}_k,F\}-x^k\; x^j\;
\{\bar{y}_k,
[\{\bar{A}_{jl},F\}\; x^l+\{\bar{B}_{jJ},F\}\; X^J
+\{\bar{C}_j^J,F\}\; Y_J]\}_{\bar{S}}
\nonumber\\
&=& 
2x^k \{\bar{y}_k,F\}-x^k\; x^j\;
[\{\bar{A}_{jk},F\}+\{\bar{B}_{jJ},F\}\; C_k^J
-\{\bar{C}_j^J,F\}\; B_{kJ}]
\nonumber\\
&=& 
2x^k [\{\bar{A}_{kj},F\} x^j+\{\bar{B}_{kJ},F\}\; X^J+\{\bar{C}_k^J,F\}\; Y_J]
-x^k\; x^j\;
[\{\bar{A}_{jk},F\}+\{\bar{B}_{jJ},F\}\; C_k^J
-\{\bar{C}_j^J,F\}\; B_{kJ}]
\nonumber\\
&=& 
x^k [\{\bar{A}_{kj},F\} x^j+2\{\bar{B}_{kJ},F\}\; X^J+2\{\bar{C}_k^J,F\}\; Y_J]
+x^k\; x^j\;
[\{\bar{C}_j^J,F\}\; B_{kJ}-\{\bar{B}_{jJ},F\}\; C_k^J]
\ea
where we have used that $A_{jk}$ is symmetrically projected.

Comparing (\ref{5.32}) and (\ref{5.49}) 
we find that to second order, modulo a term proportional to 
the constraints $y'_j$, we have $O_F=F'$ except that for $O_F$ we set 
$p=p(0)$ before evaluating the brackets $\{.,\}_S$ involving $F$ while 
for $F'$ we do not do that. Nevertheless both expressions fulfill the 
same purpose, namely they define observables with respect to 
$\tilde{C}_{j(1)}$ and $C'_{j(1)}$ respectively. The reason for this is 
clear: From the relation
\be \label{5.50}
\{C_{j(1)},C_{k(1)}\}_{\bar{S}}=\kappa_{jk(0)}\;^a\; C_{a(0)}
\ee
we find from using 
$\bar{y}_j=[(\sigma^{-1})^k y'_k]_{p=p(0)}=[(\sigma^{-1})j^k 
C_{k(0)}]_{p=p(0)}$ and the fact that restricting $p=p(0)$ does not affect
the $\{.,\}_{\bar{S}}$ bracket that $\{\bar{y}_j,\bar{y}_k\}_{\bar{S}}=0$ 
just like $\{y'_j,y'_k\}_{\bar{S}}=0$. Thus the two descriptions 
agree at second order, we are using equivalent sets of constraints. 
Note that while the constraints $\tilde{C}_j$ use $p=p(0)$ 
in their construction, the constraints
$\tilde{C_a}=O_{C_a}$ do not, the variable $p$ is not yet eliminated 
in this partial reduction approach. \\
\\
The advantage of working with the approach developed in this paper is that 
it directly generalises to higher orders. In the next section we 
develop higher order perturbation theory
for partially reduced gauge systems with backreaction which is 
directly relevant for cosmology.

\section{Partially reduced perturbation 
theory}
\label{s6}

We display here the explicit formulae for third order partially reduced 
perturbation theory, i.e. the explicit formulae for the remaining 
constraints $C_a$ in the presence of the reduction of $C_j$. Again we 
will keep the index range $a$ arbitrary but note that unless the 
range is uni-valent the subsequent perturbative constraints do not 
close at finite order (although the unperturbed constraints do).
The results of this section are directly relevant for cosmological 
perturbation theory.\\
\\
The non-perturbative expression is given by 
\be \label{6.1}
\tilde{C}_a(q,p,Q,P,X,Y):=
C_a(q,p,x=0,y=-h(q,x=0,Q,P,X,Y),Q,P,X,Y)
\ee
where $h=\sum_{n=1}^\infty h_{j(n)}$ is found using theorem \ref{th4.1}
which is a perturbative method for determining $h_j(q,x,Q,P,X,Y)$.
Conceptually, $h_j$ results from solving $C_a=0$ for 
$p_a=-\hat{h}_a(q,x,y,Q,P,X,Y)$ and solving\\ 
$C_j(q,p=-\hat{h}(q,x,y,Q,P,X,Y),x,y,Q,P,X,Y)=0$ for 
$y_j=-h_j(q,x,Q,P,X,Y)$. Although by the general theory reviewed in 
sections \ref{s3}, \ref{s5.1}
it is clear that the (\ref{6.1}) close, it is instructive 
to verify this by direct computation. To that end we introduce 
the collective configuration coordinates $k^\alpha=(q^a,Q^A,X^J)$ and 
and momentum coordinates $i_\alpha=(p_a,P_A,Y_J)$ on which $\tilde{C}_a$ 
still depends. We use the notation $F_{y=-h,x=0}$ to mean that we 
{\it first} replace $y=-h(q,x,Q,P,X,Y)$ to obtain 
$F'(q,p,x,Q,P,X,Y)=F(q,p,x,y=-h(q,x,Q,P,X,Y),Q,P,X,Y)$ and {\it then} 
set $x=0$ to obtain $\tilde{F}(q,p,Q,P,X,Y)=F'(q,p,x=0,Q,P,X,Y)$.
Direct computation yields
\ba \label{6.2}
&& 
\{\tilde{C}_a,\tilde{C}_b\}=
\frac{\partial\tilde{C}_a}{\partial i_\alpha}\;
\frac{\partial\tilde{C}_b}{\partial k^\alpha} -\; a\;\leftrightarrow \; b
\nonumber\\
&=& ([\frac{\partial C_a}{\partial i_\alpha}]_{y=-h,x=0}
-[\frac{\partial C_a}{\partial y^j}]_{y=-h,x=0}\;
[\frac{\partial h_j}{\partial i_\alpha}]_{x=0})
\;\;
([\frac{\partial C_b}{\partial k^\alpha}]_{y=-h,x=0}
-[\frac{\partial C_b}{\partial y^k}]_{y=-h,x=0}\;
[\frac{\partial h_k}{\partial k^\alpha}]_{x=0})
-\; a\;\leftrightarrow \; b
\nonumber\\
&=&
\{
\frac{\partial C_a}{\partial i_\alpha}
\;\frac{\partial C_b}{\partial k^\alpha} 
-\; a\;\leftrightarrow \; b
\}_{y=-h,x=0}
\nonumber\\
&&-
\{
[\frac{\partial C_a}{\partial y^j}]_{y=-h,x=0}\;
[\frac{\partial h_j}{\partial i_\alpha}]_{x=0}\;
[\frac{\partial C_b}{\partial k^\alpha}]_{y=-h,x=0}
+
[\frac{\partial C_a}{\partial i_\alpha}]_{y=-h,x=0}\;
[\frac{\partial C_b}{\partial y^k}]_{y=-h,x=0}\;
[\frac{\partial h_k}{\partial k^\alpha}]_{x=0}
-\; a\;\leftrightarrow \; b]_{y=-h,x=0}
\}
\nonumber\\
&& +
\{[\frac{\partial C_a}{\partial y^j} \;
\frac{\partial C_b}{\partial y^k}]_{y=-h,x=0}\;
[
\frac{\partial h_j}{\partial i_\alpha}
\frac{\partial h_k}{\partial k^\alpha}
-\; j\;\leftrightarrow \; k
]_{x=0})
\}
\nonumber\\
&=& 
\{C_a,C_b\}_{y=-h,x=0}
-\{
\frac{\partial C_a}{\partial y^j} \;
\frac{\partial C_b}{\partial x^j}
-\; a\;\leftrightarrow \; b
\}_{y=-h,x=0}
-
\{
\frac{\partial C_a}{\partial y^j}\;
\{[h_j]_{x=0},C_b\}
-\; a\;\leftrightarrow \; b
\}_{y=-h,x=0}
\nonumber\\
&& +
[\frac{\partial C_a}{\partial y^j} \;
\frac{\partial C_b}{\partial y^k}]_{y=-h,x=0}\;
\;
\{[h_j]_{x=0},[h_k]_{x=0}\}
\nonumber\\
&=& 
\{C_a,C_b\}_{y=-h,x=0}
-
\{
\frac{\partial C_a}{\partial y^j}\;
\{y_j+[h_j]_{x=0},C_b\}
-\; a\;\leftrightarrow \; b
\}_{y=-h,x=0}
\nonumber\\
&& +
[\frac{\partial C_a}{\partial y^j} \;
\frac{\partial C_b}{\partial y^k}]_{y=-h,x=0}\;
\;
\{[h_j]_{x=0},[h_k]_{x=0}\}
\nonumber\\
&=& 
\{C_a,C_b\}_{y=-h,x=0}
-
\{
\frac{\partial C_a}{\partial y^j}\;
(\{\tilde{C}_j,C_b\}+\{y_k,h_j\}\;
\frac{\partial C_b}{\partial y^k})
-\; a\;\leftrightarrow \; b
\}_{y=-h,x=0}
+
[\frac{\partial C_a}{\partial y^j} \;
\frac{\partial C_b}{\partial y^k}]_{y=-h,x=0}\;
\;
\{h_j,h_k\}_{x=0}
\nonumber\\
&=& 
\{C_a,C_b\}_{y=-h,x=0}
-
\{
\frac{\partial C_a}{\partial y^j}\;
\{\tilde{C}_j,C_b\}
-\; a\;\leftrightarrow \; b
\}_{y=-h,x=0}
\nonumber\\
&&+
[\frac{\partial C_a}{\partial y^j} \;
\frac{\partial C_b}{\partial y^k}]_{y=-h,x=0}\;
\;
\{
-\{y_k,h_j\}+\{y_j,h_k\}+\{h_j,h_k
\}_{x=0}
\}
\nonumber\\
&=& 
\{C_a,C_b\}_{y=-h,x=0}
-
\{
\frac{\partial C_a}{\partial y^j}\;
\{\tilde{C}_j,C_b\}
-\; a\;\leftrightarrow \; b
\}_{y=-h,x=0}
+
[\frac{\partial C_a}{\partial y^j} \;
\frac{\partial C_b}{\partial y^k}]_{y=-h,x=0}\;
\;
\{\tilde{C}_j,\tilde{C}_k\}_{x=0}
\nonumber\\
&=& 
[(\kappa')_{ab}\;^c \; C_c+(\kappa')_{ab}\;^j\; 
\tilde{C}_j]_{y=-h,x=0}
-
\{
\frac{\partial C_a}{\partial y^j}\;
[(\kappa')_{jb}\;^c \; C_c+(\kappa')_{jb}\;^k\; 
\tilde{C}_k]-\; a\;\leftrightarrow \; b
\}_{y=-h,x=0}
\nonumber\\
&=& 
[(\kappa')_{ab}\;^c
+2\frac{\partial C_{[a}}{\partial y_j}\;
(\kappa')_{b]j}\;^c]_{y=-h,x=0}\;\; \tilde{C}_c
=:\tilde{\kappa}_{ab}^c \; \tilde{C}_c
\ea
where in the third step we used that $[h_j]_{x=0}$ does not depend on
$x,y$, in the fourth we combined the second and third term using 
$\partial/\partial x^j(.)=\{y_j,.\}$, in the fifth we noted that 
the $y_j+[h_j]_{x=0}=[\tilde{C}_j]_{x=0}$ and that 
$\{[h_j]_{x=0},[h_k]_{x=0}\}=\{h_j,h_k\}_{x=0}$ as $h_j$ does not depend on
$y$, in the seventh and eighth we combined terms, in the nineth
we used that the set of constraints $C_a,\tilde{C}_j$ closes and whose 
structure constants were defined  in (\ref{5.1}),
in the tenth we used that $\tilde{C}_j=0$ when $y=-h$ and that 
$\tilde{C}_a=[C_a]_{y=-h,x=0}$. Note that at the reduced level 
$\tilde{C}_j\equiv 0$.\\
\\
Our task is now to expand $\tilde{C}_a$ to $N-$th order in $X,Y$. We collect 
the relevant formulae from (\ref{4.5}), (\ref{4.6})
\ba \label{6.4}
C_a &=& 
U_a+K_a^b \; p_b+L_a^j\;y_j
+A_a^{bc}\; p_b\; p_c
+B_a^{jk}\; y_j\; y_k
+C_a^{bj}\; p_b\; y_j
\nonumber\\
U_a &=& u_a
+u_a^A\; P_A+u_a^J\; Y_J
+u_a^{AB}\; P_A\;P_B+u_a^{JK}\; Y_J\; Y_K+u_a^{AJ}\; P_A\; Y_J
\nonumber\\
K_a^b &=& k_a^b+k_a^{bA}\; P_A+k_a^{bJ} Y_J
\nonumber\\
L_a^j &=& l_a^j+l_a^{jA}\; P_A+l_a^{jJ} Y_J
\ea
where the functions $A,B,C,u,k,l$ only depend on $q,x,Q,X$. We denote 
by $\bar{A},\bar{B},\bar{C},\bar{u},\bar{k},\bar{l}$ their evaluation 
at $x=0$ and by $\bar{A}_{(n)}$ etc. the homogeneous $n-$th order
contribution to its expansion with respect
to $X$. Also $\bar{U}, \bar{K}, \bar{L}$ just mean replacement of 
$u,k,l$ in $U,K,L$ by $\bar{u},\bar{k},\bar{l}$. 

We are supposed to evaluate (\ref{6.4}) at $y=-h$ and require 
the knowledge of the coefficients $h_{j(n)}$ for $n\le N$
that were constructed by the iterative scheme of theorem \ref{th4.1}. 
We denote by $\bar{h}_j$ the evaluation of $h_j(q,x,Q,P,X,Y)$ at $x=0$ and 
by $\bar{h}_{j(n)}$ the $n-$th order contribution of its expansion 
with respect to $X,Y$. Then the exact expression is 
\be \label{6.5}
\tilde{C}_a
=\bar{U}_a+\bar{K}_a^b \; p_b-\bar{L}_a^j\;\bar{h}_j
+\bar{A}_a^{bc}\; p_b\; p_c
+\bar{B}_a^{jk}\; \bar{h}_j\; \bar{h}_k
-\bar{C}_a^{bj}\; p_b\; \bar{h}_j
\ee
and the $n-$th order homogeneous contribution to its expansion with respect
to $X,Y$ is 
\be \label{6.6}
\tilde{C}_{a(n)}
=\bar{U}_{a(n)}+\bar{K}_{a(n)}^b \; p_b-
\sum_{r+s=n}\;[\bar{L}_{a(r)}^j+\bar{C}_{a(r)}^{bj}\; p_b]\;
\bar{h}_{j(s)}
+\bar{A}_{a(n)}^{bc}\; p_b\; p_c
+\sum_{r+s+t=n}\bar{B}_{a(r)}^{jk}\; \bar{h}_{j(s)}\; \bar{h}_{k(t)}
\ee
where 
\ba \label{6.7}
\bar{U}_{a(n)} &=&
\bar{u}_{a(n)}
+\bar{u}_{a(n)}^A\; P_A+\bar{u}_{a(n-1)}^J\; Y_J
+\bar{u}_{a(n)}^{AB}\; P_A\;P_B+
\bar{u}_{a(n-2)}^{JK}\; Y_J\; Y_K+\bar{u}_{a(n-1)}^{AJ} \;P_A\; Y_J
\nonumber\\
\bar{K}_{a(n)}^b &=& \bar{k}_{a(n)}^b+\bar{k}_{a(n)}^{bA}\; P_A+
\bar{k}_{a(n-1)}^{bJ} Y_J
\nonumber\\
\bar{L}_{a(n)}^j &=& \bar{l}_{a(n)}^j+\bar{l}_{a(n)}^{jA}\; P_A+
\bar{l}_{a(n-1)}^{jJ}\; Y_J
\ea
while
\be \label{6.8}
\tilde{C}_a^{(N)}=\sum_{n=0}^N\; \tilde{C}_{a(n)}
\ee
To determine the top order of $\bar{h}_{j(r)}$ needed to evaluate 
$\tilde{C}_{a(n)}$ we recall (\ref{4.9}), (\ref{4.13}) and (\ref{4.14}): 
We note that due to $h_{j(0)}\equiv 0$ also $\bar{h}_{j(0)}=0$ therefore 
the third term in (\ref{6.6}) involves only $s,t\le n-1$. Next from 
(\ref{4.9}) we see that 
$L_{a(0)}^j\equiv 0 \equiv C_{a(0)}^{bj}$ so that 
$\bar{L}_{a(0)}^j= 0 = \bar{C}_{a(0)}^{bj}$. Therefore 
also in the second term of (\ref{6.6}) only involves $s\le n-1$. 
Thus we only need $\bar{h}_{j(r)},\; r=1,..n-1$ to determine 
$\tilde{C}_{a(n)}$ which in turn requires to know 
$h_{j(r)},\; r=1,..n-1$. By (\ref{4.13}) and (\ref{4.14})  
the computation of both $h_{a(n)},h_{j(n)}$ requires knowledge of 
$h_{b(r)},\; h_{k(r)};r\le n-1$. The latter are obtained by 
solving $C_{a(r)}=C_{j(r)}=0$ for $r\le n-1$. Thus to find $h_{j(r)},\;
r\le n-1$ we need to perturbatively solve $C_{a(s)}=0,\; 0\le n-2$ and 
$C_{j(s)}=0,\;s\le n-1$ or in other words we must perturbatively solve 
$C_a^{(r)},\;r\le n-2$ and 
$C_j^{(r)},\;r\le n-1$. 
For instance to find $\tilde{C}_{a(3)}$ 
we must solve $C_{a(0)},C_{a(1)}=0$ and $C_{j(0)}=C_{j(1)}=C_{j(2)}=0$
and since $C_{a(1)}\equiv 0 \equiv C_{j(0)}$ corresponding to 
$h_{a(1)}=h_{j(0)}=0$ we just need to solve 
$C_{a(0)}=C_{j(1)}=C_{j(2)}$. It should be stressed that while the solutions 
$p_a(r)=-h_{a(r)}$ of (\ref{4.13}), (\ref{4.14}) enter the construction of 
$\tilde{C}_a$ via $h_j$, in $\tilde{C}_a$ the variable $p$ is otherwise 
still unconstrained. 
 
The object (\ref{6.8}) is to be contrasted with the strategy outlined in 
section
\ref{s2} where one expands the unperturbed $C_a,C_j$ to order $N,N-1$ 
respectively with respect to $x,y,X,Y$ relying on the assumption that 
the smearing functions $f^a,g^j$ respectively are to be considered as 
zeroth and second order quantities respectively. We compare this with 
$\tilde{C}_a^{(N)}$, which, as just derived, is obtained by 1. perturbatively 
solving the system
$C_a^{(N-2)}=C_j^{(N-1)}=0$ for $-y_j=h_j^{(N-1)}=h_{j(1)}+..+h_{j(N-1)}$, 2.
restricting to $x=0$ to find $\bar{h}_j^{(N-1)}=[h_j^{(N-1)}]_x=0$ and 3.
expanding $C_a$ evaluated at $x^j=0,y_j=-\bar{h}_j^{(N-1)}$ up to 
$N-$th order in $X,Y$. 
Common to both procedures is that both involve $C_a^{(N)},C_j^{(N-1)}$ 
but the crucial difference is that in the second procedure $C_a^{(N)}$ 
acquires the non-trivial modification $C_a^{(N)}\to \tilde{C}_a^{(N)}$ 
and forgets about $C_j^{(N-1)}$ while in the first procedure 
one keeps both $C_a^{(N)}, C_j^{(N-1)}$ and hopes that these form 
a closed system of constraints which, as we have seen in section \ref{s2},
is not the case. The modification $C_a^{(N)}\to\tilde{C}_a^{(N)}$ is just 
the perturbative counter part of switching from $C_a,C_j$ to equivalent 
constraints (i.e. they have the same kernel) $\tilde{C}_a,\tilde{C}_j$ 
for which the crucial property holds that the $\tilde{C}_j$ close among 
themselves (even Abelian) and
that the $\tilde{C}_a$ are invariant with respect to them
so that a partial reduction with respect to the $\tilde{C}_j$ can be carried 
out. 

Accordingly, $\tilde{C}_a^{(N)}$ has a profound justification 
while $C_a^{(N)\prime}$ does not. A sign of that can be seen from the 
fact that (\ref{6.8}) can be shown to close, {\it with respect to the 
full bracket} up to higher order corrections of 
order at least $N+1$. To see this, we split the exact 
$\tilde{C}_a=\tilde{C}_a^{(N)}+Z_{a(N+1)}$ where $Z_{a(N+1)}=O(N+1)$ 
contains the higher order terms of $\tilde{C}_a$ with respect to $X,Y$. 
Then using (\ref{6.2})
\ba \label{6.9}
&& \{\tilde{C}_a^{(N)},\tilde{C}_b^{(N)}\}=  
\{\tilde{C}_a-Z_{a(N+1)},
\tilde{C}_b-Z_{b(N+1)}\}
\nonumber\\
&=&\{\tilde{C}_a,\tilde{C}_b\}
-\{\tilde{C}_a,Z_{b(N+1)}\}  
+\{\tilde{C}_b,Z_{a(N+1)}\}  
+\{Z_{a(N+1)},Z_{b(N+1)}\}
\nonumber\\
&=& 
\tilde{\kappa}_{ab}\;^c\; \tilde{C}_c
-\{\tilde{C}_a,Z_{b(N+1)}\}_{\bar{S}}+O(N+1)  
+\{\tilde{C}_b,Z_{a(N+1)}\}_{\bar{S}}+O(N+1)  
+O(2N)
\nonumber\\
&=& 
\tilde{\kappa}_{ab}\;^c\; \tilde{C}_c^{(N)}
-\{\tilde{C}_{a(0)}+\tilde{C}_{a(2)}+O(3),Z_{b(N+1)}\}_{\bar{S}}
+\{\tilde{C}_{b(0)}+\tilde{C}_{b(2)}+O(3),Z_{a(N+1)}\}_{\bar{S}}
+O(N+1)
\nonumber\\
&=& 
\tilde{\kappa}_{ab}\;^c\; \tilde{C}_c^{(N)}
-\{\tilde{C}_{a(2)},Z_{b(N+1)}\}_{\bar{S}}
+\{\tilde{C}_{b(2)},Z_{a(N+1)}\}_{\bar{S}}
+O(N+1)
\nonumber\\
&=& 
\tilde{\kappa}_{ab}\;^c\; \tilde{C}_c^{(N)}
+O(N+1)
\ea
where we used that $\{.,.\}_S$ does not decrease orders, that 
$\tilde{C}_{a(0)}$ has vanishing $\{.,.\}_{\bar{S}}$ brackets and that 
$\tilde{C}_{a(1)}\equiv 0$. To see the latter we note from (\ref{6.6})
and that $h_{j(0)}=0$
\be \label{6.10}
\tilde{C}_{a(1)}
=\bar{U}_{a(1)}+\bar{K}_{a(1)}^b \; p_b-
[\bar{L}_{a(0)}^j+\bar{C}_{a(0)}^{bj}\; p_b]\;
\bar{h}_{j(1)}
+\bar{A}_{a(0)}^{bc}\; p_b\; p_c=0
\ee
since by (\ref{4.9})
\be \label{6.11}
U_{a(1)}=K_{a(1)}^b=L_{a(0)}^j=A_{a(1)}^{bc}=C_{a(0)}^{bj}=0
\ee
for all $q,x,P,Q,X,Y$ hence also at $x=0$. 

Note that the proof of (\ref{6.9}) uses the 
non-perturbative result (\ref{6.2}). 
The result (\ref{6.9}) shows that the failure of the $\tilde{C}_a^{(N)}$ 
to close exactly is of higher order O(N+1) and thus formally decays to 
zero as $N\to \infty$. It is exactly zero for any $N$ if the range 
of $a,b$ is only univalent. Since only approximately closing constraints
are not suitable for operator constraint quantisation, the partial 
reduction procedure is presumably not helpful for multivalent range of $a,b$.
See, however, \cite{Gambini} where only approximately closing constraints are 
considered as second class constraints which opens a quantisation strategy
based on second class constraints (Dirac bracket formalism). Yet, the 
result (\ref{6.9}) can be used as a motivation to follow the approach
to perturbation theory for partially reduced gauge systems with backreaction 
proposed in the present article at least in the classical 
theory. It is conceptually clear and simple and does not 
need the split of Poisson brackets into background and perturbation brackets,
no canonical transformations are involved. Yet, it can be seen as a natural
extension of the established method \cite{Gomar} as shown in the previous
section. That (\ref{6.9}) closes up to corrections of order higher than
$C_a^{(N)}$ itself is also to be contrasted with (\ref{5.39}) for 
second order which only 
closes up to the same order. 

We finish this section by examplifying the procedure 
and write $\tilde{C}_a^{(3)}$ with all details 
supplied, following the notation of theorem 
\ref{th4.1}. This requires to know $y_j(1)=-h_{j(1)}, 
y_j(2)=-h_{j(2)}$ and, as an intermediate step, 
$p_a(0)$ in the notation of 
theorem \ref{th4.1}. Here $p_a(0)$ solves $C_{a(0)}=0$ and is supposed 
to be a known function of $q,Q,P$. Following the steps of theorem 
(\ref{th4.1}) we find successively in terms 
of the zeroth order matrix 
$S_j^k(q,Q,P):=N_{j(0)}^k+2\;F_{j(0)}^{ck}\; p_c(0)$, 
which is assumed to be regular
\ba \label{6.12}
h_{k(1)} &=& (S^{-1})_k\;^j\;[V_{j(1)}+M_{j(1)}^a \; p_a(0)
+D_{j(1)}^{ab}\; p_a(0)\; p_b(0)]
\\
h_{k(2)} &=& (S^{-1})_k\;^j\;[
V_{j(2)}+M_{j(2)}^a \; p_a(0)-N_{j(1)}^k\;h_{k(1)}
+D_{j(2)}^{ab}\; p_a(0)\; p_b(0)
+E_{j(0)}^{kl}\; h_{k(1)}\; h_{l(1)}
-F_{j(1)}^{ak}\; p_a(0)\; h_{k(1)}]
\nonumber
\ea
which are to be restricted to $x=0$ thus yielding $\bar{h}_{j(1)},
\bar{h}_{j(2)}$ and inserted into (\ref{6.6}) for 
$n=2,3$ to find $\tilde{C}_{a(2)}, \tilde{C}_{a(3)}$ while 
$\tilde{C}_{a(0)}=C_{a(0)},\; \tilde{C}_{a(1)}=C_{a(1)}=0$ remain 
unmodified. Then $\tilde{C}_a^{(3)}=\sum_{n=0}^3 \tilde{C}_{a(n)}$.
Note again that $p_a(0)$ which solves 
$C_{a(0)}=0$ just enters $h_{j(1)}, h_{j(2)}$   
but the unconstrained $p$ still remains intact in all other parts 
of $\tilde{C}_{a(n)}$ in particular in $C_{a(0)}$ and 
all coefficients in $U,K,L,A,B,C$ (\ref{6.6}).

We note that in full reduction to order $N$ we require both 
$h_a,h_j$ to order $N$ and not only to order $N-2,N-1$ respectively
which then gives directly the reduced Hamiltonian 
\be \label{6.13}
H^{(N)}(t)=\dot{\tau}^a(t)\;[\sum_{n=0}^N\; h_{a(n)}]_{q=\tau,x=\rho} 
+\dot{\rho}^j(t)\;[\sum_{n=0}^N\; h_{j(n)}]_{q=\tau,x=\rho} 
\ee
subject to the gauge fixing conditions $q^a=\tau^a, x^j=\rho^j$,
as functions of $Q,P,X,Y$. Thus, to obtain for instance $H^{(3)}$ 
we require in addition $h_{a(2)}, h_{a(3)}$ if $\dot{\rho}=0$
and in addition $h_{j(3)}$ if $\dot{\rho}\not=0$ all of which can 
be obtained analogously. 

Therefore at low orders of $N$ the amount of work 
in partial reduction, when applicable, is significantly lower than for full 
reduction 
which comes at the price of having still to perform a quantum reduction of 
the $\tilde{C}_a^{(N)}$. At higher orders the difference in 
the amount of work involved becomes insignificant so that a full reduction 
seems preferrable.

\section{Conclusion and outlook}
\label{s7}

The present project was partly motivated by the desire to understand 
the seminal work \cite{Gomar} 
from a broader perspective. In \cite{Gomar}
an elegant approach to 
classical and quantum
(partially, i.e. with respect to the non-homogeneous gauge transformations)
gauge invariant second order cosmological theory including backreaction
is developed. The intention of the present study was to extract the 
symplecto-geometric origin of the methods developed in \cite{Gomar}
with the aim to generalise them to higher  orders, different symmetries 
and more general gauge systems. In fact, the generalisation of 
(cosmological) perturbation theory to higher orders is a much debated 
subject and to the best of our knowledge there is no conclusion 
in the literature how to proceed, the challenge being to define 
a perturbative notion of gauge invariance \cite{HOCPT,HOCPT1}. 

We were intrigued by the fact that 
in \cite{Gomar} sophisticated canonical transformations are employed 
in order to generate gauge invariant variables which
at least in part of its variables strongly reminded us of the first 
few terms in the Taylor expansion of relational Dirac observables 
\cite{Henneaux}. In fact, we used the relational approach before 
\cite{GHTW} but using additional matter that is designed to drastically
simplify the occurring tasks listed below and to arrive at an exact solution.

We believe that the present work at least partly 
succeeded in unveiling the broader structure 
underlying the work of \cite{Gomar}. The central ingredients are as 
follows:\\ 
1.\\ 
Starting form a physically motivated Killing symmetry, one
can naturally split the symplectic structure into a symmetric and 
a non-symmetric sector.\\
2.\\
The first class constraints of the underlying constraints also naturally 
split into ``symmetric'' and ``non-symmetric'' subsets corresponding to a
split of the tensors  smearing those constraints.\\
3.\\
One then further splits the degrees of freedom into pure gauge and true
degrees of freedom respectively. That additional split is less canonical,
the only requirement being that the gauge degrees of freedom allow 
to gauge fix the constraints. This yields altogether four sets: 
symmetric or non-symmetric gauge degrees of freedom and     
symmetric or non-symmetric true degrees of freedom. \\
4.\\
As is well known, in that situation one may pass {\it in principle} 
to the reduced phase space. This step requires solving the constraints,
the gauge fixing conditions and the stability conditions on the gauge 
fixings for the Lagrange multipliers (smearing function). For sufficiently 
complicated gauge systems, this step cannot be carried out exactly. 
The observation is (see theorem \ref{th4.1}) that precisely in the 
presence of the symmetry one may solve these three sets of equations 
perturbatively where we identify the perturbations with the non-symmetric 
degrees of freedom. Responsible for this is that the symmetry imposes that
the first order of the symmetric constraint and the zeroth order of the 
non-symmetric constraint vanish identically on the full phase space. 
This allows to establish a well defined hierarchy of equations which can 
be solved iteratively and explicitly.\\
5.\\
One may use the reduction process of 4. in two versions, the fully 
gauge invariant version and the partially gauge invariant version.
They turn out to be of similar complexity.
In the full version, all constraints, symmetric and non symmetric,
are reduced in a single step. The non-trivial task is then to 
develop the perturbation theory of the reduced (or physical) Hamiltonian.
In the partial version, only a subset of the constraints are reduced. 
These are in general not just the non-symmetric constraints because 
these generically lack the important feature of forming a Poisson subalgebra
even before perturbing them. The first non-trivial step is therefore to 
pass to an equivalent set of symmetric and non-symmetric constraints.
After the corresponding partial reduction, the remaining task is to
develop the perturbation theory for the accordingly 
reduced remaing symmetric constraints which in their exact version 
are exactly partially gauge invariant with respect to the 
non-symmetric constraints. It is the partial version that 
is the direct analog of \cite{Gomar} and when considering the second 
order of our approach we find exact match with results of \cite{Gomar}.\\
\\ 
We may apply the general theory developed in this work, among others,
to the following tasks:\\
i. Higher order cosmological perturbation theory\\
We have written out all perturbative and iterative 
formulae for a general gauge system
in sections \ref{s4}, \ref{s5} and \ref{s6} and in principle one just has 
to specialise them to the system and the order that one is interested in.
What one needs are the usual perturbative expansions of the constraints 
but its ingredients have to be combined in a non standard way in order 
that gauge invariance be maintained. 
Of particular importance is for instance third order cosmological 
perturbation theory with an eye towards non-Gaussianities \cite{NonGauss}
and the needed expressions in Hamiltonian language are already 
available \cite{E-NNT}.\\
ii. Spherically symmetric black holes\\
There is a rich literature on spherically symmetric spacetimes and 
their perturbations \cite{WZ}. 
To the best of our knowledge, this has not been done 
yet including backreaction and also only to second order which is technically
already quite challenging. If these backreaction formulae are worked 
out one would in principle be able to study quantum backreaction in 
Hawking radiation, perhaps even black hole evaporation \cite{BHE}.
As in this case the partial reduction would leave still
an infinite number of constraints, given the complications that we have 
pointed throughout the paper regarding quantisation, one would prefer 
here the fully reduced approach. We have recently started this 
programme in \cite{BHPT}. \\  
iii. Axi-symmetric black holes\\
What we have said about Schwarzschild type black holes literally also
applies to axi-symmetric (Kerr type) black holes \cite{Teukolsky}. \\
iv. Quantum Backreaction\\
This paper and most of the literature is concerned with classical 
backreaction. In \cite{ST} we have developed an approach to quantum 
backreaction based un quantum mechanical space adiabatic perturbation theory
(SAPT) \cite{SAPT} which is a perturbative expansion in addition
to that 
with respect to the non-symmetric degrees of freedom, it is based on a 
mass hierachy which typically makes the symmetric degrees of feedom 
``slower'' than the ``faster'' non-symmetric ones. 
It seems natural to extend the ideas of \cite{ST} to the black hole context.
A challenge will be to extend the Moyal product techniques of \cite{SAPT} 
to the case (black holes) that the slow sector is still a field theory 
rather than 
a quantum mechanical system as in the case of cosmology. Also, as pointed 
out in \cite{Gomar} as well as in \cite{ST}, additional canonical 
transformations on the full phase space are generically required in order 
to allow for Fock type quantisations of the non-symmetric fields because 
masses and couplings of the fast fields are generically non-constant 
functions of the slow fields which calls for Hilbert-Schmidt type 
of conditions.

}

\end{document}